\documentclass[%
 aip,
 jcp,%
 amsmath,amssymb,
 preprint,%
 groupedaddress,
]{revtex4-1}

\usepackage{graphicx}
\usepackage{multirow}
\begin{document}

\title{Molecular double core-hole electron spectroscopy for chemical analysis}

\author{Motomichi Tashiro}

\author{Masahiro Ehara}

\email{ehara@ims.ac.jp}

\affiliation{Institute for Molecular Science, Okazaki 444-8585, Japan }

\author{Hironobu Fukuzawa}

\author{Kiyoshi Ueda}

\affiliation{Institute of Multidisciplinary Research for Advanced Materials, 
Tohoku University, Sendai 980-8577, Japan }

\author{Christian Buth}

\affiliation{The PULSE Institute for Ultrafast Energy Science, SLAC National Accelerator Laboratory,
Menlo Park, California 94025, USA}

\affiliation{Department of Physics and Astronomy, Louisiana State University,  
Baton Rouge, Louisiana 70803, USA }

\author{Nikolai V. Kryzhevoi}
\author{Lorenz S. Cederbaum}

\affiliation{Theoretical Chemistry, Institute of Physical Chemistry, Heidelberg University,
69120 Heidelberg, Germany}

\date{\today}

\begin{abstract}
We explore the potential of double core hole electron spectroscopy for chemical analysis 
in terms of x-ray two-photon photoelectron spectroscopy (XTPPS).
The creation of deep single and double core vacancies induces significant reorganization of valence
electrons. The corresponding relaxation energies and the interatomic relaxation energies are
evaluated by CASSCF calculations. We propose a method how to experimentally extract these
quantities by the measurement of single and double core-hole ionization potentials (IPs and DIPs). The
influence of the chemical environment on these DIPs is also discussed for states with
two holes at the same atomic site and states with two holes at two different atomic sites. 
Electron density difference between 
the ground and double core-hole states clearly shows the relaxations accompanying the double
core-hole ionization. The effect is also compared with the sensitivity of single core hole ionization
potentials (IPs) arising in single core hole electron spectroscopy. We have demonstrated the 
method for a representative set of small molecules LiF, BeO, BF, CO, N$_2$, C$_2$H$_2$, C$_2$H$_4$,
C$_2$H$_6$, CO$_2$ and N$_2$O. The scalar relativistic effect on IPs and on DIPs are briefly 
addressed.
\end{abstract}

\maketitle

\section{Introduction}
The effect of the chemical environment manifests itself in energy differences of molecular 
core levels with respect to the atomic ones referred to as "chemical shifts". These can be measured
by core level spectroscopies, e.g., by x-ray photoelectron spectroscopy (XPS) also known as 
electron spectroscopy for chemical analysis (ESCA) and by x-ray-induced Auger electron spectroscopy 
(XAES)\cite{Siegbahn71}. Both spectroscopies have shown to be exceedingly successful tools to reveal
the quantitative elemental composition of molecules and solids. 

More than two decades ago, Cederbaum et al.\cite{Cederbaum86,Cederbaum87_1} discovered 
that the creation of double core vacancies in molecular systems probes the chemical environment 
more sensitively than the creation of single core vacancies. Two-atomic site double ionization 
potentials, or briefly two-site DIPs ( or two-site double ionization energies, DIEs ) 
are particularly sensitive to the chemical environment as  
the examples of the C$_2$H$_2$, C$_2$H$_4$, C$_2$H$_6$\cite{Cederbaum86} and 
C$_6$H$_6$\cite{Cederbaum87_1} molecules demonstrate. The chemical shifts of one-atomic site DIPs, 
or briefly one-site DIPs, were found to be similar to the chemical shifts of the single core level 
ionization potentials (IPs), or ionization energies (IEs). 
This finding has given impetus to a number of theoretical studies 
aimed at elucidating properties of molecular double core hole 
states\cite{Cederbaum87_2,Ohrendorf91,Agren93,Reynaud96}. 

So far experimental explorations of double core hole states with conventional XPS were restricted 
to those having two vacancies on the same atomic site only\cite{Kanter06,Kheifets09} since the 
probability to produce a two-site double core hole state with one-photon absorption is practically 
zero at third-generation synchrotrons due to low x-ray intensities. This prevented further progress 
of the subject. The situation has changed with development of x-ray free-electron 
lasers (x-ray FELs)\cite{Feldhaus05}. At FEL facilities in operations, such as FLASH in 
Hamburg\cite{XFEL} and SPring-8 Compact SASE Source (SCSS) test accelerator\cite{SCSS}, 
multi-photon absorption processes resulting in multiply ionized states of various systems have 
been extensively studied\cite{Sorokin06,Foehlisch07,Sato08,Jiang09,Fukuzawa09}. In the x-ray FEL 
facility LCLS at SLAC National Accelerator Laboratory, which has just started 
operations\cite{LCLS}, ultrashort pulses
of a duration about 1-fs containing 2.4$\times 10^{11}$ photons with energies of 1 keV
are expected to be generated\cite{Emma04,Galayda} thus opening up the possibility to study 
molecular two-site double core hole states. Inspired by the advent of the x-ray FEL at LCLS,
Santra et al.\cite{Santra09} have demonstrated theoretically by the proof-of-principle 
simulations on the organic para-aminophenol molecule that two-site double core hole 
states can indeed be probed by means of x-ray two-photon photoelectron spectroscopy (XTPPS).

The operating principle of XTPPS is depicted schematically in Fig.~\ref{Fig1}. The initial step in 
XTPPS corresponds conventional XPS, i.e., a neutral molecule with an energy $E_0$ is 
irradiated by an x-ray photon with an energy $\omega_X$ and a photoelectron with the kinetic energy
$\vec{\bf k}_{P,1}^2/2$ is ejected. This photoelectron carries information about a singly
core ionized state $E^+$ of the molecule. If a second x-ray photon is absorbed before the 
intermediate core hole state decays, the second photoelectron expelled from the cation with 
the kinetic energy $\vec{\bf k}_{P,2}^2/2$ carries information about a double ionization 
potential. It is important to have an intense x-ray pulse with a duration that is significantly 
shorter than the core-hole lifetimes (typical lifetimes of core ionized states of F, O, N and C
atoms are 3 to 7 femtoseconds). If the pulse duration is longer than these lifetime, then Auger 
decay is likely to occur prior to absorbing the second photon and thus the double core hole 
states may not be probed. A dicationic state $E^{++}$ of the system prepared by two-photon 
absorption decays electronically. Two primary Auger decays take place which overlap in time. 
An Auger decay happens preferably at that atomic site where the core hole has the shorter lifetime 
and an Auger electron with kinetic energy $\vec{\bf k}_{A,1}^2/2$ is ejected. This process proceeds 
in the presence of the second core hole which also decays via the Auger mechanism emitting
an electron with kinetic energy $\vec{\bf k}_{A,2}^2/2$.  The electrons ejected via such a cascade 
of Auger decays can in principle be measured by a novel Auger spectroscopy which we call x-ray 
two-photon-induced Auger electron spectroscopy (XTPAES). 

It is worthwhile to note that double core ionization can be accompanied by various shake-up 
processes similar to single core ionization. These many-body effects should manifest themselves in 
XTPPS spectra as satellites which are of interest as well. Both x-ray two-photon-induced Auger 
spectra and satellites structures will be addressed elsewhere. 
 
%In core ionization the change of geometry can be significant, depending on the case under 
%investigation. In single core ionization one can explain the measurements well by employing 
%the concept of vertical transitions. As in single ionization, also in XTPPS where the two X-ray 
%photons must be absorbed within a time shorter than the Auger decay times, the concept of 
%vertical transitions can be expected to be very useful. 

The subject of the present paper is the main double core hole states. 
In order to provide a guideline for XTPPS experiments, 
we have performed 
{\it ab initio} calculations of core level single and double ionization potentials of 
LiF, BeO, BF, CO, N$_2$, C$_2$H$_2$, C$_2$H$_4$, C$_2$H$_6$, CO$_2$, and N$_2$O molecules. 
In addition we have explored the 
sensitivity of the DIPs to the chemical environment of the core ionized atoms. 
We decompose the DIPs in three physical contributions such as the orbital energies, the electrostatic
repulsion energy between two core holes and the generalized relaxation energy and describe how
the latter can be extracted from the experimental XTPPS spectra.

\section{Computational methods}
{\it Ab initio} calculations of the vertical ionization potentials of the single and double core vacancy 
states of LiF, BeO, BF, CO, N$_2$, C$_2$H$_2$, C$_2$H$_4$, C$_2$H$_6$, CO$_2$, and N$_2$O were performed 
using the $\Delta$SCF\cite{Bagus65} and CASSCF\cite{Werner85} methods. The molecular geometries used in 
these calculations were optimized at the M{\o}ller-Plesset level of theory (MP2) \cite{Hampel92} 
employing the correlation-consistent polarized valence triple zeta (cc-pVTZ) basis sets of 
Dunning\cite{Dunning89}. Depending on whether singly or doubly ionized states were considered, 
the configurations in the CASSCF method were restricted to those having one or two holes in the 
K-shell orbitals, respectively. We used the active spaces comprising all the occupied molecular 
orbitals (except for the 1$s$ orbitals of the atoms other than H) and all the valence unoccupied 
$\pi^*$ and $\sigma^*$ ones which contain large contributions from different atomic $p$ orbitals.  
Thus, the active space of the CASSCF calculations consists of $2s$, $\sigma$, $\pi$, $\pi^*$ and 
$\sigma^*$ orbitals with core occupancy being fixed. 
The cc-pVTZ basis sets were employed in all the CASSCF and $\Delta$SCF calculations. 
For CO and C$_2$H$_2$, the cc-pVDZ, cc-pCVTZ and cc-pVQZ basis sets were also used in order to examine 
the basis set dependence of our results. 
All the single and double core-hole states were solved by independent CASSCF calculations 
using different configuration space and, therefore, the calculated states are not strictly 
orthogonal to each other. However, the one and two-site double core-hole states are well separated 
in energy and their interaction are expected to be negligible.

For molecules with equivalent atoms, N$_2$, C$_2$H$_2$, C$_2$H$_4$, C$_2$H$_6$ and CO$_2$, we calculated 
double ionization potentials using both localized and delocalized molecular orbital pictures following the
recipe given by Cederbaum et al.~\cite{Cederbaum86} and discuss differences between them.  Note that only
core orbitals were localized which was performed with the Boys method\cite{Boys66}. 
In the localized representation we obtained the ionization potentials of the one-site double core hole 
states S$_1^{-2}$ and S$_2^{-2}$ as well as of the two-site double core hole states S$_1^{-1}$S$_2^{-1}$. 
Carrying out calculations with wave functions described by the linear combinations 
S$_1^{-2}\pm$S$_2^{-2}$ gives rise to double ionization potentials in the delocalized picture. 
Differences between single ionization potentials arising due to applying localized and delocalized 
representations are not studied in the present paper because they have been discussed in detail 
before\cite{Denis75,Cederbaum77}.  

In this work, we ignore the geometry relaxation of ionized state. 
In core ionization the change of geometry can be significant, depending on the case under 
investigation. In single core ionization one can explain the measurements well by employing 
the concept of vertical transitions. As in single ionization, also in XTPPS where the two X-ray 
photons must be absorbed within a time shorter than the Auger decay times, the concept of 
vertical transitions can be expected to be very useful. 

In order to assess the impact of scalar relativistic effects on the core level single and double ionization 
potentials we made relativistic CASSCF calculations for the CO and BF molecules using the eighth order 
Douglas-Kroll-Hess Hamiltonian (DKH8)\cite{Douglas74,Hess85,Hess86,Reiher04_1,Reiher04_2}.
To get insight into the dynamic correlations, we also performed CI calculations with the CAS 
space plus single excitations from the CAS for both single and double core-hole states.

All calculations were done with the Molpro2008 quantum chemistry package\cite{MOLPRO}. 

\section{Results and discussions}
\subsection{Single core hole states}
Let us first discuss single core hole IPs.  The ionization potential for the formation of a vacancy $S^{-1}$ can be 
represented as 
\begin{eqnarray}
\label{SIP}
IP(S^{-1})=-\varepsilon_{S}-RC(S^{-1}), &&
\end{eqnarray}
where $\varepsilon_{S}$ is the corresponding orbital energy and $RC(S^{-1})$ is a contribution
to the ionization potential due to relaxation $R(S^{-1})$ and correlation $C(S^{-1})$ effects:
\begin{eqnarray}
\label{RC}
RC(S^{-1})=R(S^{-1})+C(S^{-1}) &&
\end{eqnarray}
The relaxation and electron correlations intermix with each other and cannot be 
strictly separated. The separation of these quantities was discussed in details in a perturbative way
\cite{CeDoSc1980} and in a nonperturbative way using MRCC \cite{DaMu2009}. 
The correlation contribution can be further decomposed into two parts C1 and C2 
(see Refs. \onlinecite{Pickup73} and \onlinecite{Cederbaum77}) 
where C1 describes a part of the ground state pair correlation energy 
disappearing upon removal of an electron from the spin orbital $S$, and C2 accounts for changes in 
the remaining pair correlation energy due to relaxation. Except for C1 which is a very 
small contribution, all contributions to $RC$ are thus associated with relaxation of molecular 
orbitals. Therefore, for brevity of discussion, we may call $RC$ the generalized relaxation energy. 

A straitforward way to obtain the relaxation energy is to perform $\Delta$SCF calculations. $R(S^{-1})$ 
is then derived as the difference between the respective orbital energy (with opposite sign) and the 
calculated IP. In order to get a correlation contribution to IP, post-Hartree-Fock calculations are 
generally needed. CASSCF is one of these methods. Noteworthy, in systems with core holes delocalized 
due to symmetry requirements,  C2 can be accounted by performing $\Delta$SCF calculations using 
localized orbitals instead of delocalized ones as shown by Cederbaum and Domcke\cite{Cederbaum77}.

In Table~\ref{IPs} we list IPs calculated with the CASSCF and $\Delta$SCF methods together with 
available experimental values\cite{Bakke80,Schirmer87,Kempgens97,Ehara06_1,Ehara07,Ehara06_2,Hatamoto07}. 
Table~\ref{IPs} also contains the constituting parts of IPs,
namely the orbital energies, the relaxation energies obtained from $\Delta$SCF calculations,
as well as the generalized relaxation and pure correlation contributions, both obtained from CASSCF 
calculations. The correlation contributions were calculated by subtracting the CASSCF IPs from
the $\Delta$SCF ones. Note that, since $\Delta$SCF calculations for molecules with equivalent atoms
were performed using the localized representation, the calculated IPs correspond to the localized $1s$ 
orbitals rather than to the delocalized 1$\sigma_g$ and 1$\sigma_u$.

In general, the agreement between the CASSCF and experimental results is reasonable. This concerns both
the absolute values of IPs and the g-u energy splittings for molecules with equivalent atoms. 
Except for basis set effects which always are an issue in {\it ab initio} calculations, and relativistic 
effects, deviations from the experiment are attributed to the lack of dynamic correlations in the ground 
and single core hole states, and to the core-valence separation approximation employed in the calculations. 
We notice that influences of the above-mentioned effects and approximations partially compensate for each 
other. Indeed, performing calculations without core-valence separability lowers IPs\cite{Angonoa87}. 
A lowering of IPs can also be achieved by improving basis sets. On the other hand, taking into account 
relativistic effects increases IPs. In Appendices~\ref{app1} and \ref{app2} we explore the basis sets and 
relativistic effects in more detail.

It is interesting to compare the different contributions to IPs in Eq. (\ref{SIP}). After the orbital energy, the 
relaxation energy represents the largest constituent part of a core level IP. It increases nearly 
proportional to the atomic number $Z$. For some molecules, however, remarkable 
deviations from this trend occur under influence of the chemical environment. A crucial role 
for the relaxation energy plays the change of the electron density distribution of valence electrons, 
$\Delta \rho$, in an atom due to a formation of chemical bonds with neighbors, and the interaction of 
$\Delta \rho$ with the core hole. Ionic bonds give rise to the strongest changes of the electron 
density distribution. As a consequence, the relaxation energies associated with core ionization of  
electron acceptors in ionic molecules (e.g.\ O and F in BeO and LiF, respectively) are noticeably 
larger than the relaxation energies of the same atoms bound by covalent bonds with their neighbors 
(O and F in CO and BF, respectively). For other factors influencing the relaxation energies see 
Ref. \onlinecite{Cederbaum86}. 

In comparison to relaxation effects, correlation effects induced by core ionization are rather small.
According to our calculations, the magnitude of the static correlation effects does not exceed 3 eV for 
the molecules studied and accounting for missing dynamic correlation can hardly modify this situation 
dramatically. Interestingly, the largest correlation effects manifest themselves in atoms whose 
neighbors are the strong electron acceptors O and F. 

The effect of the chemical environment on core level ionization potentials of various systems including the 
molecules explored here is rather well established and we therefore refrain from long discussions
on this subject. We only mention that the chemical environment is able to introduce large changes
in the ionization potentials as, for example, can be realized by comparing molecules with ionic and 
covalent bonds. On the other hand, in the sequence of the C$_2$H$_2$, C$_2$H$_4$, C$_2$H$_6$ molecules
characterized by the triple, double and single carbon-carbon bond, respectively, the impact of the 
chemical environment is rather moderate. In contrast to single core hole ionization potentials, double 
core hole ionization potentials reveal much more pronounced sensitivity to the chemical environment as 
it was first demonstrated in Refs. \onlinecite{Cederbaum86} and \onlinecite{Cederbaum87_1}.

\subsection{Double core hole states}
\subsubsection{General equations and results}
In analogy to Eq.~(\ref{SIP}), we represent the double ionization potential of a state
with two core vacancies $S_i^{-1}$ and $S_j^{-1}$ as
\begin{eqnarray}
\label{DIP}
DIP(S_i^{-1}, S_j^{-1})=-\varepsilon_{S_i}-\varepsilon_{S_j}-RC(S_i^{-1},S_j^{-1})+E_{RE}, &&
\end{eqnarray}
where $E_{RE}$ is the repulsion-exchange energy of the two core holes. For an one-site double core 
hole state, it is described by the two-electron integral $V_{S_iS_iS_iS_i}$, or $(S_iS_i|S_iS_i)$, and, 
for a two-site double core hole state, by a linear combination of the integrals $V_{S_iS_jS_iS_j}$ 
and $V_{S_iS_jS_jS_i}$ 
where the exchange term is negligibly small when the core holes are well localized\cite{Ohrendorf91}.

The generalized relaxation $RC(S_i^{-1},S_j^{-1})$ can be decomposed into three parts
\begin{eqnarray}
\label{IRC}
RC(S_i^{-1},S_j^{-1})=RC(S_i^{-1})+RC(S_j^{-1})+NRC(S_i^{-1},S_j^{-1}), &&
\end{eqnarray}
where $RC(S_i^{-1})$ given by Eq.~(\ref{RC}) describes relaxation and correlation effects 
induced by creation of the core vacancy $S_i^{-1}$ as there were no core vacancy $S_j^{-1}$. 
The relaxation and correlation energies are expected to be non-additive upon creation of multiple 
vacancies. A possible deviation from additivity is thus described in Eq.~(\ref{IRC}) by the 
{\it non-additivity} term $NRC(S_i^{-1},S_j^{-1})$. 

Depending on whether two core holes were created on the same atomic site or on different atomic sites,
$NRC(S_i^{-1},S_j^{-1})$ may be called the {\it excess} generalized relaxation energy, 
$ERC(S_i^{-1},S_i^{-1})$,  or  the {\it interatomic} generalized relaxation energy, 
$IRC(S_i^{-1},S_j^{-1})$.  Note that, while the $RC(S_i^{-1})$ and $ERC(S_i^{-1},S_i^{-1})$
measure local properties of a core ionized atom, $IRC(S_i^{-1},S_i^{-1})$ measures the impact of the 
environment "between" the atoms involved.

In Table~\ref{DIPs} we list the calculated double ionization potentials. We also show the correlation
contributions to DIPs. As one can see these contributions are remarkably larger than those to the single 
IPs and may constitute 5.6 eV. In the special cases, however, when we performed calculations with delocalized 
core orbitals, differences between $\Delta$SCF and CASSCF values rise to 27-35 eV resulting from the failure 
of the $\Delta$SCF method in the delocalized picture to account for all relaxation contributions as 
described by Cederbaum et al.\cite{Cederbaum86,Cederbaum87_1}. 

One can notice by comparing Table~\ref{IPs} and ~\ref{DIPs} that the impact of the chemical environment 
is different for double and single ionization potentials. Of particular interest is to compare 
two-site DIPs with single IPs since their sensitivities to the chemical environment reveal major
differences. A prominent example already discussed in detail in Ref.~\onlinecite{Cederbaum86} is the 
hydrocarbons C$_2$H$_2$, C$_2$H$_4$ and C$_2$H$_6$. Here, the chemical shifts in the two-site DIPs are 
much more pronounced than in the single IPs being also attributed to different carbon-carbon bondlengths 
resulted from a different number of hydrogen atoms in these molecules. In XPS one can hardly distinguish 
between these three compounds while in XTPPS this should be possible in principle. The situation is somewhat
different for the individual molecule N$_2$O.  In this molecule the sensitivity of two-site DIPs to the 
chemical environment
is lower compared to that of single IPs. Indeed, the core ionization potentials of the terminal
and central nitrogen atoms differ by 4 eV whereas the difference between the N$_t$1s$^{-1}$O1s$^{-1}$ 
and N$_c$1s$^{-1}$O1s$^{-1}$ double ionization potentials constitutes 2.3-2.8 eV. The latter energy 
difference is much lower than 11 eV which one would expect taking into account only the differences between
the N$_c$O and  N$_t$O bondlengths and between the single core hole ionization potentials. As we show below, 
the reason for such a dramatic reduction of the chemical shift has to do with distinct relaxation processes 
induced by the creation of different pairs of core holes.

Taking into account Eqs.~(\ref{SIP}) and (\ref{IRC}), we can represent $DIP(S_i^{-1}, S_j^{-1})$ as 
\begin{eqnarray}
\label{DIP.IP}
DIP(S_i^{-1}, S_j^{-1})=IP(S_i^{-1})+IP(S_j^{-1})-NRC(S_i^{-1},S_j^{-1})+E_{RE}, &&
\end{eqnarray}
and define the ionization potential of the core vacancy $S_i^{-1}$ in the presence
of the core vacancy $S_j^{-1}$ as 
\begin{eqnarray}
\label{IP.Si}
IP(S_j^{-1}; S_i^{-1}) = IP(S_i^{-1})-NRC(S_i^{-1},S_j^{-1})+E_{RE},&&
\end{eqnarray}
whereas 
\begin{eqnarray}
\label{IP.Sj}
IP(S_i^{-1}; S_j^{-1}) = IP(S_j^{-1})-NRC(S_i^{-1},S_j^{-1})+E_{RE}&&
\end{eqnarray}
is defined as the ionization potential of the core vacancy $S_j^{-1}$ in the presence of the core vacancy 
$S_i^{-1}$.

Both the ionization potentials $IP(S_i^{-1})$ of a neutral system and the ionization potentials 
$IP(S_i^{-1}; S_j^{-1})$  of a core-ionized one can be obtained experimentally, e.g., by means of XTPPS. 
In XTPPS, the kinetic energy $KE(S_i^{-1})$ of the first photoelectron ejected from the orbital $S_i$ 
defines $IP(S_i^{-1})$, whereas the kinetic energy $KE(S_i^{-1}; S_j^{-1})$ of the second photoelectron 
ejected from the orbital $S_j$ defines $IP(S_i^{-1}; S_j^{-1})$. Obviously, the sum 
$IP(S_i^{-1})+IP(S_i^{-1}; S_j^{-1})$ gives $DIP(S_i^{-1}, S_j^{-1})$. As shown below, important 
properties of the system under study can be extracted also from the measurable energy difference
\begin{eqnarray}
\label{dE}
\Delta E &=& KE(S_i^{-1})-KE(S_j^{-1}; S_i^{-1})= IP(S_j^{-1}; S_i^{-1})- IP(S_i^{-1})   \nonumber   \\
         &=& DIP(S_j^{-1}, S_i^{-1})-IP(S_i^{-1})-IP(S_j^{-1}).
\end{eqnarray}

Similar to $DIP(S_i^{-1}, S_j^{-1})$, the kinetic energy $KE(S_i^{-1}; S_j^{-1})$ depends 
significantly on the mutual arrangement of the core vacancies $S_j^{-1}$ and $S_i^{-1}$ in a 
molecule. This is clearly seen from Table~\ref{CO} where we collect the kinetic energies of 
all the core electrons of the CO molecule which one would detect in an XTPPS experiment 
given that the molecule is irradiated by an x-ray pulse with photon energies of 1 keV.
First of all, we notice that it is more difficult to remove an electron from the core ionized CO
molecule than from the neutral one. The respective energy difference is about 70-90 eV when 
the first and the second core electrons are ejected from the same core orbital. 
This energy difference reduces drastically to about 15 eV when different core orbitals are
affected. Apparently, the electrostatic interaction between the two core holes plays a
crucial role here. $NRC(S_i^{-1}, S_j^{-1})$ exerts an influence on the above energy 
differences too, as can be deduced from Eqs.~(\ref{IP.Si}) and (\ref{IP.Sj}).

\subsubsection{One-site double core hole states}
If $S_i=S_j=S$ then $\Delta E$ takes the form
\begin{eqnarray}
\label{dE1}
 \Delta E1(S^{-2})=-ERC(S^{-1},S^{-1}) + V_{SSSS}. 
\end{eqnarray}
We calculated $\Delta E1(S^{-2})$ for the molecules under study using the respective CASSCF single and 
double core hole ionization potentials and collect them in Table~\ref{deltaEs}. The dependence of 
$\Delta E1(S^{-2})$ on the atomic number $Z$ is displayed in Fig.~\ref{Fig2}(a).

The excess generalized relaxation energy $ERC(S^{-1},S^{-1})$ can be easily obtained by  
measuring the energy difference $\Delta E1(S^{-2})$ provided that the integral $V_{SSSS}$ is known. 
$V_{SSSS}$ can be extracted from {\it ab initio} Hartree-Fock calculations on the electronic ground
state of neutral molecules. Alternatively, it can be calculated by using the approximate analytical 
expression suggested in Ref. \onlinecite{Cederbaum86} : 
\begin{eqnarray}
\label{Vcccc}
V_{SSSS}=(2^{5/2}/3\pi)(Z-2^{-3/2}).
\end{eqnarray}
The respective results for $V_{SSSS}$ and a difference between them are discussed in Appendix~\ref{app3}.

Using Eq.~(\ref{IRC}) we represent $ERC(S^{-1},S^{-1})$ as 
\begin{eqnarray}
\label{ERC}
ERC(S^{-1},S^{-1})=RC(S^{-1},S^{-1})-2RC(S^{-1}). &&
\end{eqnarray}
It has been shown in Ref. \onlinecite{Cederbaum86} that at the second order perturbation theory
the following relationship between the relaxation energies is valid:
\begin{eqnarray}
\label{R.ratio}
R(S^{-1},S^{-1})= 4 R(S^{-1}).
\end{eqnarray}
Since the impact of correlation into ionization potentials is small compared to the impact 
of relaxation, we expect that a similar relationship exists between the generalized relaxation
energies $RC(S^{-1},S^{-1})$ and $RC(S^{-1})$.   Let us therefore introduce that 
\begin{eqnarray}
\label{RC.ratio}
RC(S^{-1},S^{-1})= n\times RC(S^{-1}),
\end{eqnarray}
and find the optimal $n$. After the substitution of Eqs.~(\ref{ERC}) and (\ref{RC.ratio}) 
into (\ref{dE1}), we get
\begin{eqnarray}
\label{dE1.n}
 \Delta E1(S^{-2})=-(n-2)RC(S^{-1}) + V_{SSSS}.
\end{eqnarray}
Now we can easily calculate $n$ by using the {\it ab initio} results for $\Delta E1$, $RC$ and $V_{SSSS}$.  
The respective values of $n$ as a function of the atomic number $Z$ are shown in Fig. \ref{Fig3}. As one can 
see, deviations of the calculated $n$ from the expected value of 4 are rather small (15\% in the worst case 
of Li) and therefore $n=4$ can be considered as a plausible approximation for the molecules studied in 
the present work. 

As a result, we obtain the following expression for the generalized relaxation energy:
\begin{eqnarray}
\label{RC.1site}
RC(S^{-1})= (V_{SSSS}- \Delta E1(S^{-2}))/2.
\end{eqnarray}
The values of $RC(S^{-1})$ calculated by means of Eq.~(\ref{RC.1site}) are given in Table~\ref{deltaEs} 
and also plotted in Fig. \ref{Fig2}(b) as a function of $Z$. It is worthwhile to note a reasonable agreement 
between them and the corresponding results from Table ~\ref{IPs}.

\subsubsection{Two-site double core hole states}
If $S_i \neq S_j$  then $\Delta E$ takes the form
\begin{eqnarray}
\label{dE2}
 \Delta E2(S_i^{-1},S_j^{-1})=-IRC(S_i^{-1},S_j^{-1}) + 1/r,
\end{eqnarray}
where the repulsion-exchange energy $E_{RE}$ has been approximated by the inverse  of the distance
$r$ between the two atoms with the core vacancies $S_i^{-1}$ and $S_j^{-1}$. We calculated 
$\Delta E2(S_i^{-1},S_j^{-1})$ using the CASSCF double and single core hole ionization potentials
and collected these results in Table ~\ref{deltaEs}. 

By looking at Eq.~(\ref{dE2}), one expects a linear dependence between $\Delta E2(S_i^{-1},S_j^{-1})$ 
and $1/r$ .  This expectation 
is not realized however as seen from Fig. \ref{Fig4}(a) where a variation of $\Delta E2$ with $1/r$ is shown. 
To elucidate the complex behavior of $\Delta E2$ we calculated the interatomic generalized relaxation 
energy $IRC(S_i^{-1},S_j^{-1})$ from Eq.~(\ref{dE2}) and plotted these results as a function 
of $r$ in Fig. \ref{Fig4}(b).  We found both positive and negative values of $IRC$ (see also Table ~\ref{deltaEs})
which indicate on an enhancement or suppression of relaxation effects, respectively.
  
In the case of diatomic molecules, $IRC$ is always negative and thus the relaxation is suppressed. 
This suppression may be interpreted by considering the change of the electron density. A core hole 
$S_i^{-1}$ attracts valence electrons and increases the electron density in its vicinity, yielding 
a deficiency of the electron density in the vicinity of the atom with a core orbital $S_j$. 
The relaxation energy corresponding to the creation of the core vacancy $S_j^{-1}$  in the presence
of the core hole  $S_i^{-1}$ is therefore reduced. The amplitudes of  $IRC$ are smaller for LiF, BF, 
and N$_2$ than for BeO, and CO. This is because the electrons are strongly attracted to the F site in
LiF and BF, or tightened in the triple bond in N$_2$ and thus the electron density flow due to core hole
creation is suppressed by these chemical environments.

Values of $IRC(S_i^{-1},S_j^{-1})$ for the polyatomic molecules C$_2$H$_2$, C$_2$H$_4$, and C$_2$H$_6$ are,
in contrast, positive. The enhancement of the relaxation for these molecules occurs due to 
the electron density flowing from the C-H bonds to the two carbon core hole sites.

The interatomic generalized relaxation energy exhibits a very interesting behavior with $r$ in triatomic molecules 
as seen from Fig. \ref{Fig4}(b). $IRC(S_i^{-1},S_j^{-1})$ is positive for CO$_2$ and N$_2$O when $S_i^{-1}$ and 
$S_j^{-1}$ are in adjacent atoms, namely in C and O in CO$_2$ and in N$_t$ and N$_c$ or in N$_c$ and O 
in  N$_t$N$_c$O. In these cases, the third atom plays the role of an electron donor and enhances the 
relaxation of the double core hole in the other two sites. The values of $IRC$ are, on the other hand, 
negative for CO$_2$ and N$_2$O with two holes in the terminal atoms, namely with holes in the two O sites 
of CO$_2$ and with holes in the N$_t$ and O sites of N$_2$O. In these cases, the creation of the core hole 
on one site already withdraws the electron density from the central atom and thus reduces the possibility 
of relaxation due to the creation of the second hole in the other terminal site. 

In order to analyze the reorganization caused by double core hole ionization, we calculated the
electron density difference between the ground and double core hole ionized states. 
These electron density differences 
without the 1s contribution are plotted in Fig. \ref{Fig5}, to better visualize the reorganization of the valence
electrons. In the blue or green region, the electron density increases, while the density decreases in
the red region. In the C$_11s^{-2}$ state of C$_2$H$_4$, the electron density of the C$_1$-C$_2$, 
C$_1$-H bond and H atoms connected to C$_1$ atom reduces and flows to the region around C$_1$ atom as 
shown in Fig. \ref{Fig5}(a). In the C$_11s^{-1}$C$_21s^{-1}$ state of C$_2$H$_4$, on the other hand, the electron 
density in the region of C-H bonds and H atoms flows to the region of both C atoms as in Fig. \ref{Fig5}(b). This 
explains the positive value of $IRC$ in the C$_11s^{-1}$C$_21s^{-1}$ state as noted above. In the case of 
the O$1s^{-2}$ state of CO, the electron density of the CO bond is used for the reorganization around the 
O atom (Fig. \ref{Fig5}(c)).

\section {Conclusions}
We have computed the ionization potentials of single and double core hole states of the small molecules 
LiF, BeO, BF, CO, N$_2$, C$_2$H$_2$, C$_2$H$_4$, C$_2$H$_6$, CO$_2$, and N$_2$O by means of the $\Delta$SCF 
and CASSCF methods in order to explore the impact of the chemical environment on the respective ionization
processes and provide a guidance for x-ray two-photon photoelectron spectroscopy (XTPPS) experiments. 

Our calculations have demonstrated that except for N$_2$O, the double ionization potentials, especially the
two-site ones of all these molecules are more sensitive to the chemical environment than the single ionization 
potentials. The sensitivity to the chemical environment of the two-site DIPs of N$_2$O is mainly governed
by the interatomic relaxation energies which, in turn, strongly depend on the electron density distribution
between the core-ionized atoms. 

The quantities extracted from XTPPS are differences between the kinetic energies of core electrons ejected via 
the first and second ionization steps, i.e., of core electrons ejected from neutral and core-ionized systems, 
respectively.  These kinetic energy differences are defined by a localization of the two core vacancies created 
and by relaxation processes induced by double core ionization.
We have shown how one can extract the generalized relaxation energy associated with single core
ionization as well as the excess and interatomic generalized relaxation energies associated with one-site
and two-site double core ionizations, respectively, from experimental data by knowing the repulsion energy 
between the two core holes.  The corresponding XTPPS experiments are now in preparation in the x-ray free
electron laser facility LCLS at SLAC National Accelerator Laboratory.

\begin{acknowledgements}
M.E. acknowledges the support from JST-CREST and a Grant-in-Aid for Scientific Research from the
Japan Society for the Promotion of Science, the Next Generation Supercomputing Project, and the
Molecular-Based New Computational Science Program, NINS. H.F. and K.U. acknowledge the support 
for the X-ray Free Electron Laser Utilization Research Project of Ministry of Education, Culture, 
Sports, Science and Technology of Japan (MEXT). C.B. was supported by National Science Foundation
under the grants No PHY-0701372 and No PHY-0449235. The Heidelberg group is grateful for financial 
support by the Deutsche Forschungsgemeinschaft. 
The computations were partly performed using the Research Center for 
Computational Science, Okazaki, Japan. 
\end{acknowledgements}

\appendix
\section{Scalar relativistic effects}
\label{app1}
It is essential in view of their large impact to take into account relativistic effects when 
considering systems with heavy atoms. For light atoms, relativistic effects are of less importance
but still should be accounted when highly accurate results are needed. In the present work we assess    
the impact of scalar relativistic effects on IPs and DIPs by carrying out calculations with the relativistic 
DKH8 Hamiltonian for the CO and BF molecules. The results of these calculations are shown in 
Table ~\ref{RelCorr}. As one can see the magnitude of the scalar relativistic effects on the single
ionization potentials grows with the increasing atomic mass constituting 0.03, 0.09, 0.35, and 0.59 eV
in the case of the B, C, O and F atoms, respectively. One of us has shown  that a similar tendency is 
observed also for the third-row Si, P, S, Cl atoms\cite{Ehara09}. A growth of scalar relativistic effects
with the atomic mass exhibits also in the case of double core hole ionization. We note that the 
relativistic effects on the one-site DIPs are about 2.3-3 times larger than those on the respective single 
IPs. Interestingly that in the case of two-site doubly ionized states the relativistic effects 
are perfectly described by the sum of the relativistic effects associated with the constituting single 
core vacancies. 

\section{Basis set effects}
\label{app2}
In this section we explore the basis set dependence of the single and double core hole ionization 
potentials by examples of the C$_2$H$_2$ and CO molecules. We have examined four basis sets:
the Dunning's correlation-consistent basis sets cc-pVXZ (X=D,T,Q) and the cc-pCVTZ one. The latter 
contains tight basis functions which are added for a better description of properties
of core-level states. The results are collected in Table~\ref{BasSet}. By looking at the sequence of 
the cc-pVXZ results we find significant differences
between the cc-pVDZ and cc-pVQZ values, especially in the case of the one-site DIPs where differences
of nearly 10 eV are present. The major changes occur upon improving the basis set from a double- to 
triple-zeta quality. Choosing the cc-pVQZ basis sets lowers ionization potentials by only 0.2-1.2 eV 
relative to the cc-pVTZ values. The cc-pCVTZ results are lower in energy than
the cc-pVTZ ones whereby they nearly coincide with the cc-pVQZ results in the case of the CO molecule. 
We used the cc-pVTZ basis throughout as a compromise between the accuracy of the results
and the computational costs. For the double core-hole states, the relaxation of valence orbitals is 
important, in particular for the one-site states where all 1s electrons on one atom are missing.

\section{Evaluation of the $V_{SSSS}$ integral}
\label{app3}
The approximate analytical expression (\ref{Vcccc}) was proposed in Ref. \onlinecite{Cederbaum86} for the two electron 
integral $V_{SSSS}$. By comparing the analytical results with the explicit {\it ab initio} ones (dotted curve and 
filled circles in Fig. \ref{Fig6}, respectively) we found a progressively growing deviation between them with
increasing $Z$ (1.5 eV for lithium, 4.5 eV for fluorine). This deviation can be removed by substituting 
1.037$\cdot$Z instead of Z in Eq.~(\ref{Vcccc}), where the coefficient 1.037 has been extracted from a 
linear fit (dashed curve in Fig. \ref{Fig6}) of the {\it ab initio} results.

\section{Effect of dynamic correlations}
\label{app4} 
We performed the CI calculations with the space of the singly excited configurations from the CASSCF
configurations and examined the semi-internal correlation.
The results for the single and double core-hole states of C$_2$H$_2$ and CO were summarized in 
Table \ref{Corr}. The difference between the results of CI and CASSCF provides the effect of the semi-internal correlation. The
semi-internal correlation has small effect on IPs of the single hole states, less than +0.03 eV for C1s$^{-1}$ 
and +0.25 eV for O1s$^{-1}$. The effect for the one-site double core-hole states is significant as +0.24 $\sim$ 
+0.54 eV, while it is small for the two-site double core-hole states as +0.08 $\sim$ +0.12 eV.

\clearpage
\newpage

%$$$$$$$$$$$$$$$$$$$$$$$$$$$$$$$$$$$$$$$$$$$$$$$$$$$$$$$$$$$$$$$$$$$$
%\begin{table}[h!]
\begin{table}
\caption{\label{IPs} Single core hole ionization potentials  and their constituent parts (in eV) as
calculated with the $\Delta$SCF and CASSCF methods. Experimental values of the ionization
potentials, where available, taken from Refs. 38-44 %\onlinecite{Bakke80,Schirmer87,Kempgens97,Ehara06_1,Ehara07,Ehara06_2,Hatamoto07}
are also shown. The cc-pVTZ basis sets were employed. } 
\begin{tabular}{ccccccccc}
\hline
\hline
\multirow{2}{*}{Molecule} & Core level,   & \multirow{2}{*}{\ -$\varepsilon_S $ \ } & 
\multicolumn{2}{c}{$\Delta$SCF} & \multicolumn{3}{c}{CASSCF} & \multirow{2}{*}{Expt.} \\
\cline{4-5} \cline{6-8}
 &  $S$ & & IP & $R(S^{-1})$  & IP & $RC(S^{-1})$ & $C(S^{-1})$ \\
\hline
LiF        & Li1s &  66.407 &  65.460 &  0.947 &  65.334 &  1.073 & 0.126 &  \\
           & F1s  &\ 710.484\ &\ 688.187\ &\ 22.297\ &\ 688.044\ &\ 22.440\ &\ 0.143\ &  \\
BeO        & Be1s & 128.383 & 124.406 &  3.977 & 123.399 &  4.984 & 1.007 &  \\
           & O1s  & 556.694 & 535.181 & 21.513 & 535.075 & 21.619 & 0.106 &  \\
BF         & B1s  & 209.735 & 202.539 &  7.196 & 201.724 &  8.011 & 0.815 &  \\
           & F1s  & 717.810 & 695.873 & 21.937 & 695.915 & 21.895 &-0.042 &  \\
CO         & C1s  & 309.111 & 298.256 & 10.855 & 296.358 & 12.753 & 1.898 & 296.069 \\
           & O1s  & 562.348 & 542.801 & 19.547 & 542.820 & 19.528 &-0.019 & 542.543 \\
N$_2$O  & N$_t$1s & 427.159 & 409.615 & 17.544 & 408.614 & 18.545 & 1.001 & 408.5 \\
        & N$_c$1s & 432.005 & 415.373 & 16.632 & 412.524 & 19.481 & 2.849 & 412.5 \\
           & O1s  & 563.760 & 543.046 & 20.714 & 542.537 & 21.223 & 0.509 & 542.0 \\
CO$_2$     & C1s  & 311.930 & 300.607 & 11.323 & 297.647 & 14.283 & 2.960 & 296.78 \\
           & O1g  & 561.956 & \multirow{2}{*}{541.979} & 19.977 & 542.870 & 19.086 &-0.891 & \multirow{2}{*}{540.6}  \\
           & O1u  & 561.954 &                          & 19.975 & 542.868 & 19.086 &-0.889 &  \\
C$_2$H$_2$ & C1g  & 305.897 & \multirow{2}{*}{292.535} & 13.362 & 292.202 & 14.062 & 0.700 & \multirow{2}{*}{291.14, 291.20}  \\ 
           & C1u  & 305.794 &                          & 13.259 & 292.111 & 14.054 & 0.795 &  \\
C$_2$H$_4$ & C1g  & 305.557 & \multirow{2}{*}{291.801} & 13.756 & 291.344 & 14.213 & 0.457 & \multirow{2}{*}{290.70, 290.88}  \\
           & C1u  & 305.508 &                          & 13.707 & 291.297 & 14.211 & 0.504 &  \\                  
C$_2$H$_6$ & C1g  & 305.040 & \multirow{2}{*}{291.774} & 13.266 & 291.147 & 13.893 & 0.627 & \multirow{2}{*}{290.76, 290.74}  \\ 
           & C1u  & 305.023 &                          & 13.249 & 291.125 & 13.898 & 0.649 &  \\
N$_2$      & N1g  & 426.686 & \multirow{2}{*}{411.242} & 15.444 & 411.027 & 15.659 & 0.215 & 409.93 \\
           & N1u  & 426.588 &                          & 15.346 & 410.932 & 15.656 & 0.310 & 409.82 \\
\hline
\hline
\end{tabular}
\end{table}
%$$$$$$$$$$$$$$$$$$$$$$$$$$$$$$$$$$$$$$$$$$$$$$$$$$$$$$$$$$$$$$$$$$$$

\clearpage
\newpage

%$$$$$$$$$$$$$$$$$$$$$$$$$$$$$$$$$$$$$$$$$$$$$$$$$$$$$$$$$$$$$$$$$$$$
%\begin{table}[h!]
\begin{table}
\caption{\label{DIPs} Calculated double core hole ionization potentials and the static 
correlation energies C (in eV). T and S refer to triplet and singlet couplings of two
core holes created on different atomic sites, respectively.
The cc-pVTZ basis sets were employed. 
}
\begin{tabular}{clrrr}
\hline
\hline
\multirow{2}{*}{Molecule} & \multirow{2}{*}{Core level} & $\Delta$SCF & \multicolumn{2}{c}{CASSCF}  \\
\cline{4-5}
 & & DIP & DIP & C  \\
\hline
LiF        & Li1s$^{-2}$ &  173.125 &  172.595 &  0.530 \\
           & F1s$^{-2}$  & 1480.418 & 1481.495 & -1.077 \\
           & Li1s$^{-1}$F1s$^{-1}$, S  & 763.447  & 763.211 & 0.236 \\
           & Li1s$^{-1}$F1s$^{-1}$, T  & 763.443  & 763.277 & 0.166 \\
BeO        & Be1s$^{-2}$ &  300.585 &  298.032 & 2.553  \\
           & O1s$^{-2}$  & 1158.135 & 1159.351 & -1.216 \\
           & Be1s$^{-1}$O1s$^{-1}$, S  & 672.823  & 671.801 & 1.022 \\
           & Be1s$^{-1}$O1s$^{-1}$, T  & 672.823  & 672.128 & 0.695 \\
BF         & B1s$^{-2}$  &  468.243 &  465.323 & 2.920 \\
           & F1s$^{-2}$  & 1494.930 & 1495.809 & -0.879 \\
           & B1s$^{-1}$F1s$^{-1}$, S   & 910.946  & 910.568 & 0.378 \\
           & B1s$^{-1}$F1s$^{-1}$, T   & 910.946  & 910.678 & 0.268 \\
CO         & C1s$^{-2}$  &  667.902 &  664.418 &  3.484 \\
           & O1s$^{-2}$  & 1175.376 & 1176.561 & -1.185 \\
           & C1s$^{-1}$O1s$^{-1}$, S   & 857.072  & 854.743 & 2.329 \\
           & C1s$^{-1}$O1s$^{-1}$, T   & 857.072  & 855.200 & 1.872 \\
N$_2$O     & N$_t$1s$^{-2}$ &  894.485  &  893.926  & 0.559  \\
           & N$_c$1s$^{-2}$ &  906.773  &  902.312  & 4.461  \\
           & O1s$^{-2}$     & 1173.683  & 1173.249  & 0.434  \\
           & N$_t$1s$^{-1}$N$_c$1s$^{-1}$, S  & 838.282 & 832.962 & 5.320 \\
           & N$_t$1s$^{-1}$N$_c$1s$^{-1}$, T  & 838.279 & 833.215 & 5.064 \\
           & N$_t$1s$^{-1}$O1s$^{-1}$, S  & 965.806 & 963.041 & 2.765  \\
           & N$_t$1s$^{-1}$O1s$^{-1}$, T  & 965.806 & 963.266 & 2.540  \\
           & N$_c$1s$^{-1}$O1s$^{-1}$, S  & 968.082 & 965.793 & 2.289  \\
           & N$_c$1s$^{-1}$O1s$^{-1}$, T  & 968.082 & 965.623 & 2.459  \\
\end{tabular} 
\end{table}
%$$$$$$$$$$$$$$$$$$$$$$$$$$$$$$$$$$$$$$$$$$$$$$$$$$$$$$$$$$$$$$$$$$$$

\clearpage
\newpage

\begin{center}
\begin{tabular}{clrrr}
CO$_2$     & C1s$^{-2}$  &  670.280 &  664.629 & 5.651  \\
           & O1s$^{-2}$  & 1172.821 & 1171.909 & 0.912  \\ 
           & C1s$^{-1}$O1s$^{-1}$, S  & 854.682 & 851.059 & 3.623  \\
           & C1s$^{-1}$O1s$^{-1}$, T  & 854.682 & 851.199 & 3.483  \\
           & O$_1$1s$^{-1}$O$_2$1s$^{-1}$, S  & 1094.795 & 1094.090 & 0.705  \\
           & O$_1$1s$^{-1}$O$_2$1s$^{-1}$, T  & 1094.795 & 1094.167 & 0.628  \\
C$_2$H$_2$ & C$_1$1s$^{-2}$  & 651.265 & 650.228 & 1.037 \\
           & C$_2$1s$^{-2}$  & 651.265 & 650.228 & 1.037 \\
           & C$_1$1s$^{-1}$C$_2$1s$^{-1}$, S  & 598.281 & 594.590 & 3.691 \\
           & C$_1$1s$^{-1}$C$_2$1s$^{-1}$, T  & 598.281 & 595.197 & 3.084 \\
           & C$_1$1s$^{-2}$-C$_2$1s$^{-2}$    & 681.646 & 651.334 & 30.312 \\
           & C$_1$1s$^{-2}$+C$_2$1s$^{-2}$    & 681.646 & 651.334 & 30.312 \\
C$_2$H$_4$ & C$_1$1s$^{-2}$  & 648.964 & 648.556 & 0.408 \\
           & C$_2$1s$^{-2}$  & 648.964 & 648.556 & 0.408 \\
           & C$_1$1s$^{-1}$C$_2$1s$^{-1}$, S  & 594.850 & 591.514 & 3.336 \\
           & C$_1$1s$^{-1}$C$_2$1s$^{-1}$, T  & 594.850 & 591.956 & 2.894 \\
           & C$_1$1s$^{-2}$-C$_2$1s$^{-2}$    & 679.386 & 649.703 & 29.683 \\
           & C$_1$1s$^{-2}$+C$_2$1s$^{-2}$    & 679.386 & 649.703 & 29.683 \\
C$_2$H$_6$ & C$_1$1s$^{-2}$  & 649.714 & 648.827 & 0.887 \\
           & C$_2$1s$^{-2}$  & 649.714 & 648.827 & 0.887 \\
           & C$_1$1s$^{-1}$C$_2$1s$^{-1}$, S  & 591.447 & 589.007 & 2.440 \\
           & C$_1$1s$^{-1}$C$_2$1s$^{-1}$, T  & 591.447 & 589.192 & 2.255 \\
           & C$_1$1s$^{-2}$-C$_2$1s$^{-2}$    & 677.339 & 649.898 & 27.441 \\
           & C$_1$1s$^{-2}$+C$_2$1s$^{-2}$    & 677.339 & 649.898 & 27.441 \\
N$_2$      & N$_1$1s$^{-2}$  & 901.704 & 901.155 & 0.549  \\
           & N$_2$1s$^{-2}$  & 901.704 & 901.155 & 0.549  \\
           & N$_1$1s$^{-1}$N$_2$1s$^{-1}$, S  & 839.912 & 835.784 & 4.128  \\
           & N$_1$1s$^{-1}$N$_2$1s$^{-1}$, T  & 839.912 & 836.437 & 3.475  \\
           & N$_1$1s$^{-2}$-N$_2$1s$^{-2}$    & 938.943 & 903.727 & 35.216 \\
           & N$_1$1s$^{-2}$+N$_2$1s$^{-2}$    & 938.943 & 903.727 & 35.216 \\
\hline
\hline
\end{tabular}
\end{center} 

\clearpage
\newpage

%$$$$$$$$$$$$$$$$$$$$$$$$$$$$$$$$$$$$$$$$$$$$$$$$$$$$$$$$$$$$$$$$$$$$
%\begin{table}[h!]
\begin{table}
\caption{\label{CO} Kinetic energies of photoelectrons one would detect in an
XTPPS experiment by irradiating the CO molecule with an x-ray pulse with photon energies of
1 keV. The sequence of the core vacancies in the records reflects the sequence of ionization steps. 
T and S refer to triplet and singlet couplings of two core holes created on different atomic 
sites, respectively. }
\begin{tabular}{lc}
\hline
\hline
\ \ \ \ \ \ \ State   & KE (eV)  \\ 
\hline
C1s$^{-1}$   &\ \ 703.642 \ \ \\
C1s$^{-1}$C1s$^{-1}$     & 631.940 \\
O1s$^{-1}$C1s$^{-1}$, S  & 688.077 \\
O1s$^{-1}$C1s$^{-1}$  T  & 687.620 \\
\hline
O1s$^{-1}$   & 457.180 \\
O1s$^{-1}$O1s$^{-1}$  & 366.259 \\
C1s$^{-1}$O1s$^{-1}$, S & 441.615 \\
C1s$^{-1}$O1s$^{-1}$, T & 441.518 \\
\hline
\hline
\end{tabular}
\end{table}
%$$$$$$$$$$$$$$$$$$$$$$$$$$$$$$$$$$$$$$$$$$$$$$$$$$$$$$$$$$$$$$$$$$$$

%$$$$$$$$$$$$$$$$$$$$$$$$$$$$$$$$$$$$$$$$$$$$$$$$$$$$$$$$$$$$$$$$$$$$
%\begin{table}[h!]
\begin{table}
\caption{\label{deltaEs} Calculated energy differences $\Delta E1(S_i^{-2})$ and  
$\Delta E2(S_i^{-1},S_j^{-1})$ together with the intra- and interatomic generalized relaxation 
energies $RC(S_i^{-1})$ and $IRC(S_i^{-1},S_j^{-1})$ (in eV).}
\begin{tabular}{clrlr}
\hline
\hline
Molecule&\multicolumn{2}{c}{Energy difference}&\multicolumn{2}{c}{Generalized relaxation energy }\\
\hline
LiF        & $\Delta E1$(Li1s$^{-2}$)              &  41.927 &  $RC$(Li1s$^{-1}$)             &  1.41  \\
           & $\Delta E1$(F1s$^{-2}$)               & 105.407 &  $RC$(F1s$^{-1}$)              & 20.17  \\
           & $\Delta E2$(Li1s$^{-1}$, F1s$^{-1}$)  &   9.833 & $IRC$(Li1s$^{-1}$, F1s$^{-1}$) & -0.74  \\
BeO        & $\Delta E1$(Be1s$^{-2}$)              &  51.234 &  $RC$(Be1s$^{-1}$)             &  5.23  \\
           & $\Delta E1$(O1s$^{-2}$)               &  89.201 &  $RC$(O1s$^{-1}$)              & 19.90  \\
           & $\Delta E2$(Be1s$^{-1}$, O1s$^{-1}$)  &  13.327 & $IRC$(Be1s$^{-1}$, O1s$^{-1}$) & -2.69  \\
BF         & $\Delta E1$(B1s$^{-2}$)               &  61.875 &  $RC$(B1s$^{-1}$)              &  8.38  \\
           & $\Delta E1$(F1s$^{-2}$)               & 103.979 &  $RC$(F1s$^{-1}$)              & 20.88  \\ 
           & $\Delta E2$(B1s$^{-1}$, F1s$^{-1}$)   &  12.929 & $IRC$(B1s$^{-1}$, F1s$^{-1}$)  & -1.52  \\
CO         & $\Delta E1$(C1s$^{-2}$)               &  71.702 &  $RC$(C1s$^{-1}$)              & 11.87  \\
           & $\Delta E1$(O1s$^{-2}$)               &  90.921 &  $RC$(O1s$^{-1}$)              & 19.03  \\ 
           & $\Delta E2$(C1s$^{-1}$, O1s$^{-1}$)   &  15.565 & $IRC$(C1s$^{-1}$, O1s$^{-1}$)  & -2.80  \\
N$_2$      & $\Delta E1$(N1s$^{-2}$)               &  79.196 &  $RC$(N1s$^{-1}$)              & 17.65  \\ 
           & $\Delta E2$(N$_1$1s$^{-1}$, N$_2$1s$^{-1}$)   &  13.825 & $IRC$(N$_1$1s$^{-1}$, N$_2$1s$^{-1}$)  & -0.65  \\
C$_2$H$_2$ & $\Delta E1$(C1s$^{-2}$)               &  66.653 &  $RC$(C1s$^{-1}$)              & 15.06  \\ 
           & $\Delta E2$(C$_1$1s$^{-1}$, C$_2$1s$^{-1}$)   &  11.015 & $IRC$(C$_1$1s$^{-1}$, C$_2$1s$^{-1}$)  & 0.96  \\ 
C$_2$H$_4$ & $\Delta E1$(C1s$^{-2}$)               &  65.915 &  $RC$(C1s$^{-1}$)              & 15.42  \\ 
           & $\Delta E2$(C$_1$1s$^{-1}$, C$_2$1s$^{-1}$)   &   8.873 & $IRC$(C$_1$1s$^{-1}$, C$_2$1s$^{-1}$)  & 1.94  \\                    
C$_2$H$_6$ & $\Delta E1$(C1s$^{-2}$)               &  66.555 &  $RC$(C1s$^{-1}$)              & 15.06  \\ 
           & $\Delta E2$(C$_1$1s$^{-1}$, C$_2$1s$^{-1}$)   &   6.735 & $IRC$(C$_1$1s$^{-1}$, C$_2$1s$^{-1}$)  & 2.72  \\                    
CO$_2$     & $\Delta E1$(C1s$^{-2}$)               &  69.335 &  $RC$(C1s$^{-1}$)              & 13.76  \\ 
           & $\Delta E1$(O1s$^{-2}$)               &  86.171 &  $RC$(O1s$^{-1}$)              & 22.70  \\ 
           & $\Delta E2$(C1s$^{-1}$, O1s$^{-1}$)   &  10.572 & $IRC$(C1s$^{-1}$, O1s$^{-1}$)  &  1.79  \\ 
           & $\Delta E2$(O1s$^{-1}$, O1s$^{-1}$)   &   8.352 & $IRC$(O1s$^{-1}$, O1s$^{-1}$)  & -2.19  \\ 
N$_2$O     & $\Delta E1$(N$_t$1s$^{-2}$)           &  76.698 &  $RC$(N$_t$1s$^{-1}$)          & 17.78  \\ 
           & $\Delta E1$(N$_c$1s$^{-2}$)           &  77.264 &  $RC$(N$_c$1s$^{-1}$)          & 17.47  \\ 
           & $\Delta E1$(O1s$^{-2}$)               &  88.175 &  $RC$(O1s$^{-1}$)              & 20.40  \\ 
           & $\Delta E2$(N$_t$1s$^{-1}$, N$_c$1s$^{-1}$)   &  11.827 & $IRC$(N$_t$1s$^{-1}$, N$_c$1s$^{-1}$)  & 1.11  \\                    
           & $\Delta E2$(N$_t$1s$^{-1}$, O1s$^{-1}$)       &  10.732 & $IRC$(N$_t$1s$^{-1}$, O1s$^{-1}$)      & 0.07  \\                    
           & $\Delta E2$(N$_c$1s$^{-1}$, O1s$^{-1}$)       &  11.890 & $IRC$(N$_c$1s$^{-1}$, O1s$^{-1}$)      &-6.00  \\                    
\hline
\hline
\end{tabular}
\end{table}
%$$$$$$$$$$$$$$$$$$$$$$$$$$$$$$$$$$$$$$$$$$$$$$$$$$$$$$$$$$$$$$$$$$$$

%$$$$$$$$$$$$$$$$$$$$$$$$$$$$$$$$$$$$$$$$$$$$$$$$$$$$$$$$$$$$$$$$$$$$
%\begin{table}[h!]
\begin{table}
\caption{\label{RelCorr}  Single and double core hole ionization potentials
of CO and BF calculated with the CASSCF method using the relativistic DKH8 Hamiltonian
and the differences between them and the respective non-relativistic results given in
Tables \ref{IPs} and \ref{DIPs}. All energies are in eV. T and S refer 
to triplet and singlet couplings of two core holes created on different atomic sites, respectively.}
\begin{tabular}{cccc}
\hline
\hline
Molecule&State&Ionization Potential&Difference\\
\hline
CO&C$1s^{-1}$& 296.446 & 0.088 \\
  &O$1s^{-1}$& 543.166 & 0.346 \\
  &C$1s^{-2}$& 664.661 & 0.243 \\
  &O$1s^{-2}$&1177.408 & 0.847 \\
  &C$1s^{-1}$O$1s^{-1}$, S& 855.176 & 0.433 \\
  &C$1s^{-1}$O$1s^{-1}$, T& 855.632 & 0.432 \\
BF&B$1s^{-1}$& 201.758 & 0.034 \\
  &F$1s^{-1}$& 696.500 & 0.585 \\
  &B$1s^{-2}$& 465.425 & 0.102 \\
  &F$1s^{-2}$&1497.177 & 1.368 \\
  &B$1s^{-1}$F$1s^{-1}$, S& 911.187 & 0.619 \\
  &B$1s^{-1}$F$1s^{-1}$, T& 911.297 & 0.619 \\
\hline
\hline
\end{tabular}
\end{table}
%$$$$$$$$$$$$$$$$$$$$$$$$$$$$$$$$$$$$$$$$$$$$$$$$$$$$$$$$$$$$$$$$$$$$

%$$$$$$$$$$$$$$$$$$$$$$$$$$$$$$$$$$$$$$$$$$$$$$$$$$$$$$$$$$$$$$$$$$$$
%\begin{table}[h!]
\begin{table}
\caption{\label{BasSet}  Basis set dependence of the CASSCF single and double core hole ionization
potentials. All energies are in eV. T and S refer 
to triplet and singlet couplings of two core holes created on different atomic sites, respectively.}
\begin{tabular}{cccccc}
\hline
\hline
Molecule&State& cc-pVDZ & cc-pVTZ & cc-pCVTZ & cc-pVQZ \\
\hline
C$_2$H$_2$ & C$1g^{-1}$& 293.262 & 292.202 & 292.111 & 291.728 \\
           & C$1u^{-1}$& 293.172 & 292.111 & 292.020 & 291.634 \\
           & C$1s^{-2}$& 658.167 & 650.228 & 649.412 & 649.711 \\
           & C$_11s^{-1}$C$_21s^{-1}$, S& 598.350 & 594.590 & 594.337 & 593.908 \\
           & C$_11s^{-1}$C$_21s^{-1}$, T& 598.853 & 595.197 & 594.971 & 594.592 \\
CO         & C$1s^{-1}$& 298.062 & 296.358 & 296.239 & 296.229 \\
           & O$1s^{-1}$& 544.881 & 542.820 & 542.611 & 542.559 \\
           & C$1s^{-2}$& 672.908 & 664.418 & 663.663 & 663.632 \\
           & O$1s^{-2}$& 1184.846 & 1176.561 & 1175.469 & 1175.373 \\
           & C$1s^{-1}$O$1s^{-1}$, S& 859.146 & 854.743 & 854.452 & 854.425   \\
           & C$1s^{-1}$O$1s^{-1}$, T& 859.491 & 855.200 & 854.931 & 854.905   \\
\hline
\hline
\end{tabular}
\end{table}
%$$$$$$$$$$$$$$$$$$$$$$$$$$$$$$$$$$$$$$$$$$$$$$$$$$$$$$$$$$$$$$$$$$$$

\clearpage
\newpage

%$$$$$$$$$$$$$$$$$$$$$$$$$$$$$$$$$$$$$$$$$$$$$$$$$$$$$$$$$$$$$$$$$$$$
%\begin{table}[h!]
\begin{table}
\caption{\label{Corr}  Effect of dynamic correlations examined by the CI calculation.  
All energies are in eV. }
\begin{tabular}{cccc}
\hline
\hline
Molecule&State& CASSCF & CI \\
\hline
C$_2$H$_2$ & C$1g^{-1}$& 292.202 & 292.216 \\
           & C$1u^{-1}$& 292.111 & 292.127 \\
           & C$1s^{-2}$& 650.228 & 650.586 \\
           & C$_11s^{-1}$C$_21s^{-1}$, S& 594.590 & 594.728 \\
           & C$_11s^{-1}$C$_21s^{-1}$, T& 595.197 & 595.319 \\
CO         & C$1s^{-1}$& 296.358  & 296.386  \\
           & O$1s^{-1}$& 542.820  & 543.070  \\
           & C$1s^{-2}$& 664.418  & 664.658  \\
           & O$1s^{-2}$& 1176.561 & 1177.096 \\
           & C$1s^{-1}$O$1s^{-1}$, S& 854.743 & 854.833  \\
           & C$1s^{-1}$O$1s^{-1}$, T& 855.200 & 855.276  \\
\hline
\hline
\end{tabular}
\end{table}
%$$$$$$$$$$$$$$$$$$$$$$$$$$$$$$$$$$$$$$$$$$$$$$$$$$$$$$$$$$$$$$$$$$$$

\clearpage
\newpage

%$$$$$$$$$$$$$$$$$$$$$$$$$$$$$$$$$$$$$$$$$$$$$$$$$$$$$$$$$$$$$$$$$$$$
\centerline{FIGURE CAPTIONS}
%$$$$$$$$$$$$$$$$$$$$$$$$$$$$$$$$$$$$$$$$$$$$$$$$$$$$$$$$$$$$$$$$$$$$
\begin{figure}
\caption{\label{Fig1} (Color online). Schematic picture of x-ray two-photon photoelectron 
spectroscopy (XTPPS) and x-ray two-photon-induced Auger electron spectroscopy (XTPAES). 
See text for explanations. In this picture it is assumed that the second photon is absorbed
before Auger decay takes place and that one core hole decays much faster than the other. In
reality all processes overlap.}
\end{figure}
%$$$$$$$$$$$$$$$$$$$$$$$$$$$$$$$$$$$$$$$$$$$$$$$$$$$$$$$$$$$$$$$$$$$$

%$$$$$$$$$$$$$$$$$$$$$$$$$$$$$$$$$$$$$$$$$$$$$$$$$$$$$$$$$$$$$$$$$$$$
\begin{figure}
\caption{\label{Fig2} (a) The energy difference $\Delta E1(S^{-2})=DIP(S^{-1},S^{-1})-[IP(S^{-1})+IP(S^{-1})]$ 
as a function of the atomic number Z of the atom with the core orbital $S$;
(b) The generalized relaxation energy $RC(S^{-1})$ calculated by means of Eq.~(\ref{RC.1site})
as a function of Z.}
\end{figure}
%$$$$$$$$$$$$$$$$$$$$$$$$$$$$$$$$$$$$$$$$$$$$$$$$$$$$$$$$$$$$$$$$$$$$

%$$$$$$$$$$$$$$$$$$$$$$$$$$$$$$$$$$$$$$$$$$$$$$$$$$$$$$$$$$$$$$$$$$$$
\begin{figure}
\caption{\label{Fig3} The ratio $n=RC(S^{-1},S^{-1})/RC(S^{-1})$ as a function of the atomic number Z 
of the atom with the core orbital $S$.  }
\end{figure}
%$$$$$$$$$$$$$$$$$$$$$$$$$$$$$$$$$$$$$$$$$$$$$$$$$$$$$$$$$$$$$$$$$$$$

%$$$$$$$$$$$$$$$$$$$$$$$$$$$$$$$$$$$$$$$$$$$$$$$$$$$$$$$$$$$$$$$$$$$$
\begin{figure}
\caption{\label{Fig4} (a) The energy difference $\Delta E2(S_i^{-1},S_j^{-1})=DIP(S_i^{-1},S_j^{-1})-[IP(S_i^{-1})+IP(S_j^{-1})]$ 
as a function of the inverse distance $r$ between the atoms with the core orbitals $S_i$ and $S_j$;
(b) The interatomic generalized relaxation energy $IRC(S_i^{-1},S_j^{-1})$ calculated by means of Eq.~(\ref{dE2})
as a function of $r$.}
\end{figure}
%$$$$$$$$$$$$$$$$$$$$$$$$$$$$$$$$$$$$$$$$$$$$$$$$$$$$$$$$$$$$$$$$$$$$

%$$$$$$$$$$$$$$$$$$$$$$$$$$$$$$$$$$$$$$$$$$$$$$$$$$$$$$$$$$$$$$$$$$$$
\begin{figure}
\caption{\label{Fig5} Differences between the valence electron density distributions  
of the ground and double core hole states: (a) C$_1$1s$^{-2}$ of C$_2$H$_4$; 
(b) C$_1$1s$^{-1}$C$_2$1s$^{-1}$ (singlet) of C$_2$H$_4$; (c) O1s$^{-2}$ of CO }
\end{figure}
%$$$$$$$$$$$$$$$$$$$$$$$$$$$$$$$$$$$$$$$$$$$$$$$$$$$$$$$$$$$$$$$$$$$$

%$$$$$$$$$$$$$$$$$$$$$$$$$$$$$$$$$$$$$$$$$$$$$$$$$$$$$$$$$$$$$$$$$$$$
\begin{figure}
\caption{\label{Fig6} The two-electron integral $V_{SSSS}$ as a function of the atomic number Z 
of the atom with the core orbital $S$. The results extracted from the {\it ab initio}
calculations (filled circles) are compared with those calculated by means of Eq.~(\ref{Vcccc})
(dotted curve). A linear fit of the {\it ab initio} results is also shown (dashed curve). }
\end{figure}
%$$$$$$$$$$$$$$$$$$$$$$$$$$$$$$$$$$$$$$$$$$$$$$$$$$$$$$$$$$$$$$$$$$$$

\clearpage
\newpage

\thispagestyle{empty}
%$$$$$$$$$$$$$$$$$$$$$$$$$$$$$$$$$$$$$$$$$$$$$$$$$$$$$$$$$$$$$$$$$$$$
\begin{figure}
\begin{center}
\scalebox {0.7} {\includegraphics{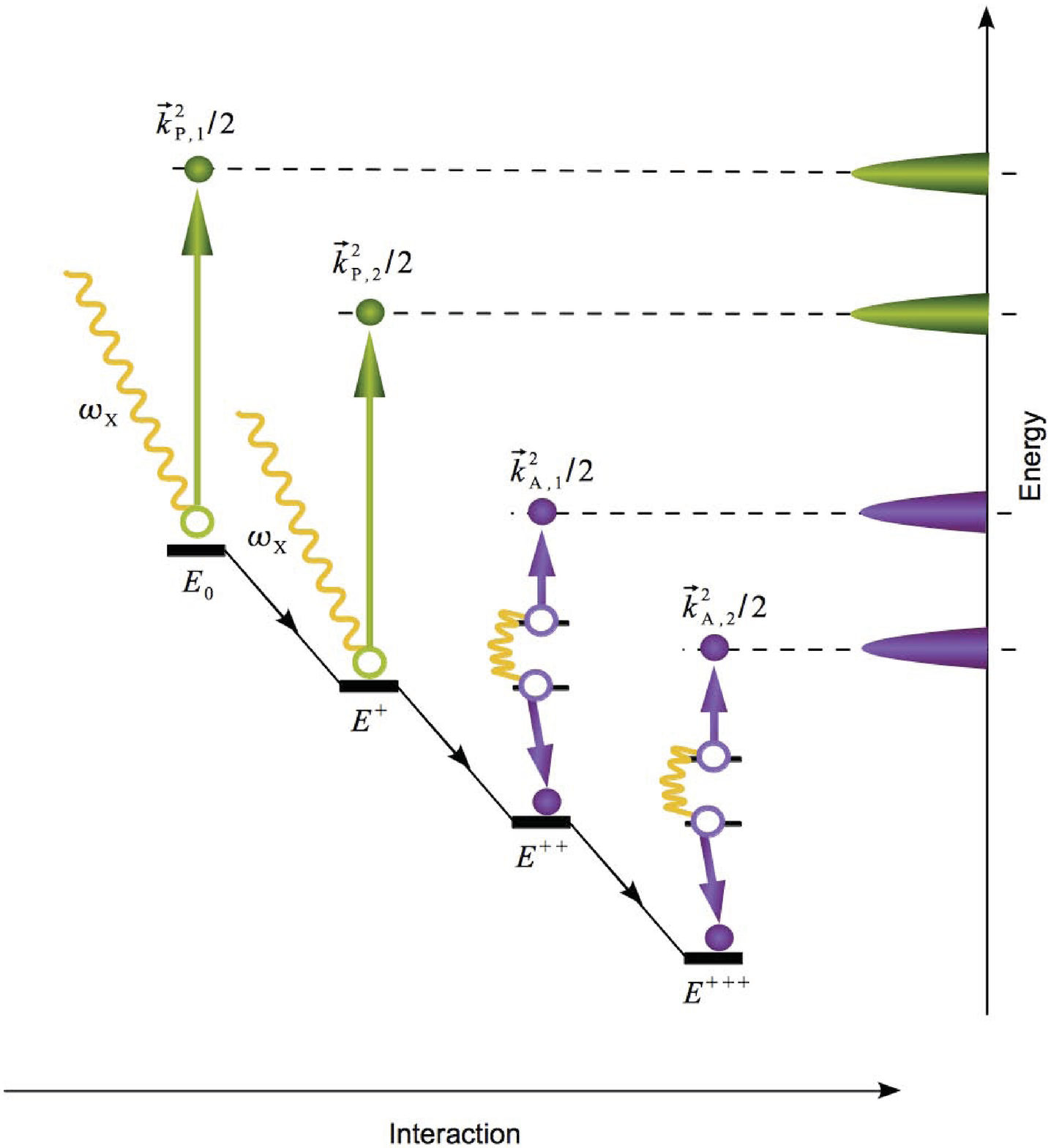}}
\end{center}
\end{figure}
%$$$$$$$$$$$$$$$$$$$$$$$$$$$$$$$$$$$$$$$$$$$$$$$$$$$$$$$$$$$$$$$$$$$$

\clearpage
\newpage
%\thispagestyle{empty}
%$$$$$$$$$$$$$$$$$$$$$$$$$$$$$$$$$$$$$$$$$$$$$$$$$$$$$$$$$$$$$$$$$$$$
\begin{figure}
\begin{center}
\includegraphics[scale=0.5]{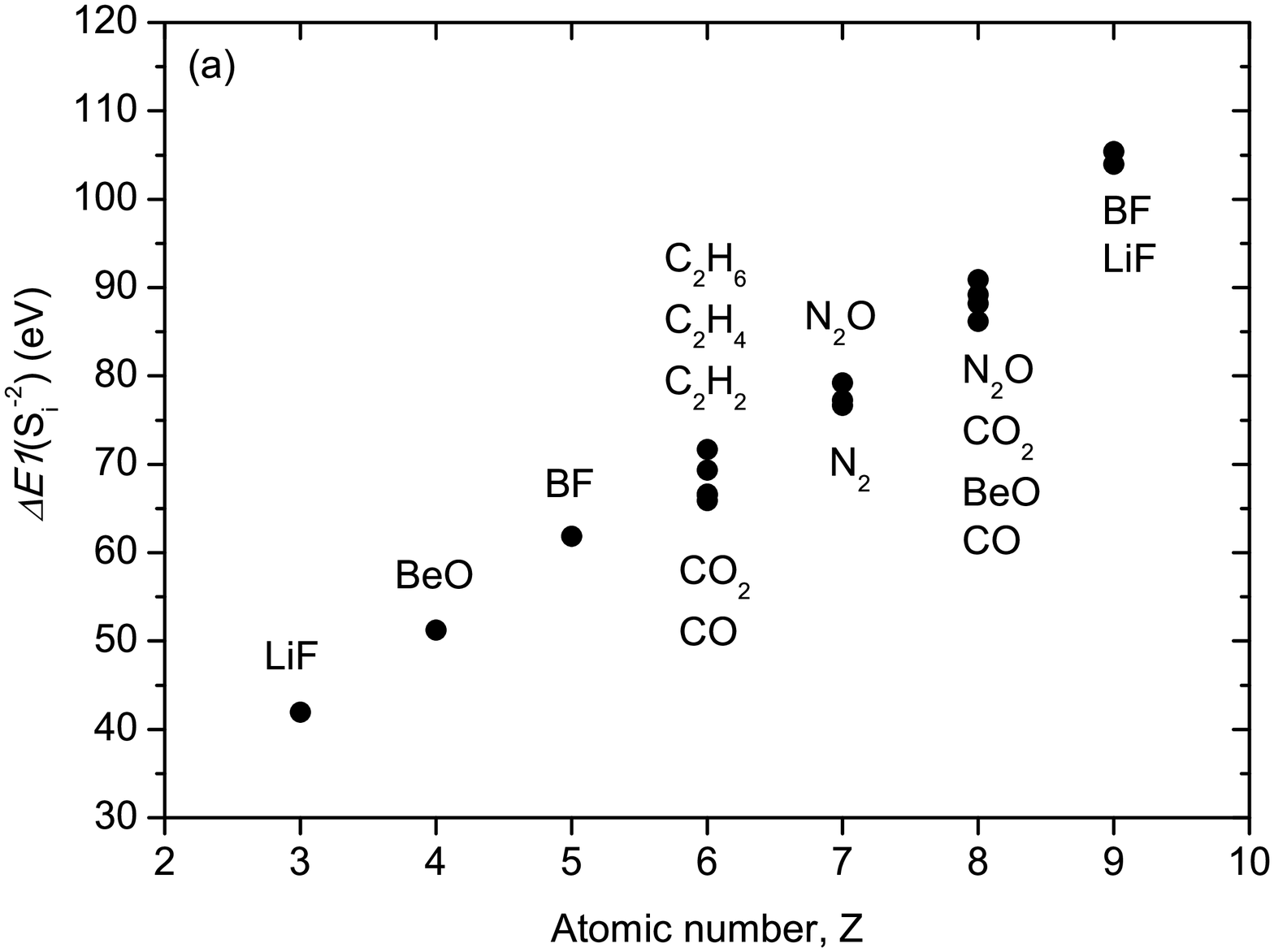}
\includegraphics[scale=0.5]{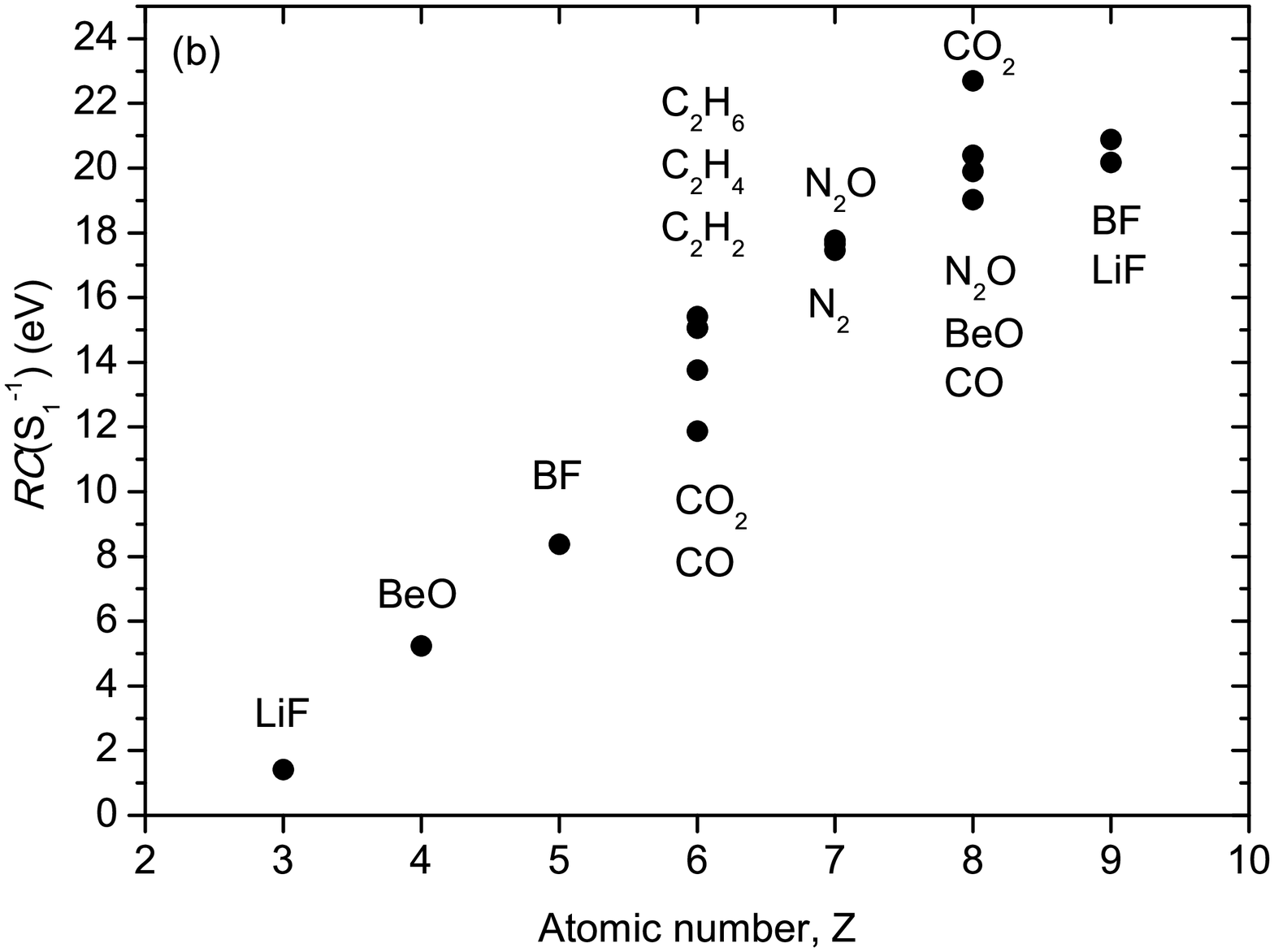}
\end{center}
\end{figure}
%$$$$$$$$$$$$$$$$$$$$$$$$$$$$$$$$$$$$$$$$$$$$$$$$$$$$$$$$$$$$$$$$$$$$

\clearpage
\newpage
\thispagestyle{empty}
%$$$$$$$$$$$$$$$$$$$$$$$$$$$$$$$$$$$$$$$$$$$$$$$$$$$$$$$$$$$$$$$$$$$$
\begin{figure}
\begin{center}
\scalebox {0.88} {\includegraphics{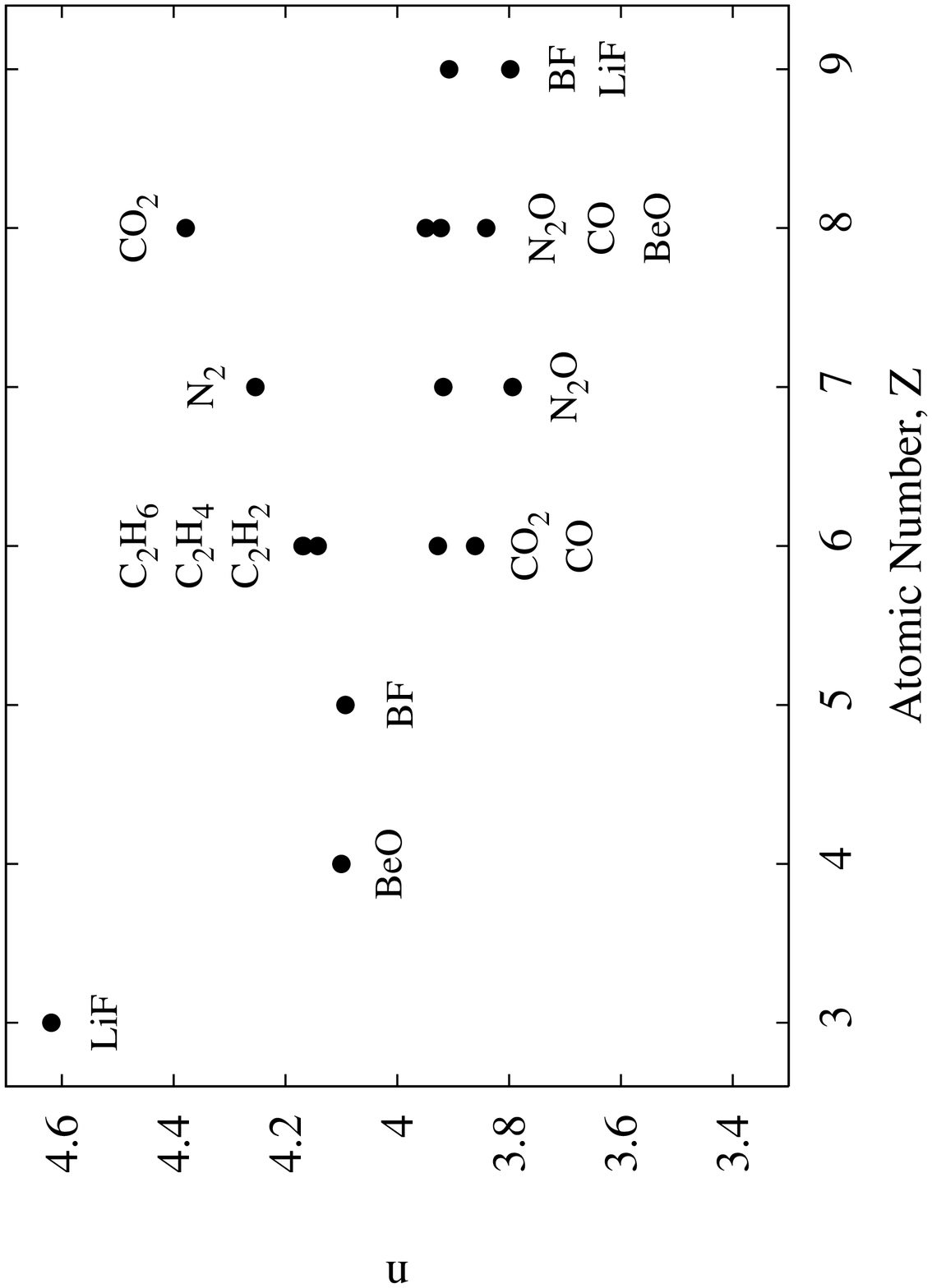}}
\end{center}
\end{figure}
%$$$$$$$$$$$$$$$$$$$$$$$$$$$$$$$$$$$$$$$$$$$$$$$$$$$$$$$$$$$$$$$$$$$$

\clearpage
\newpage
%\thispagestyle{empty}
%$$$$$$$$$$$$$$$$$$$$$$$$$$$$$$$$$$$$$$$$$$$$$$$$$$$$$$$$$$$$$$$$$$$$
\begin{figure}
\begin{center}
\includegraphics[scale=0.5]{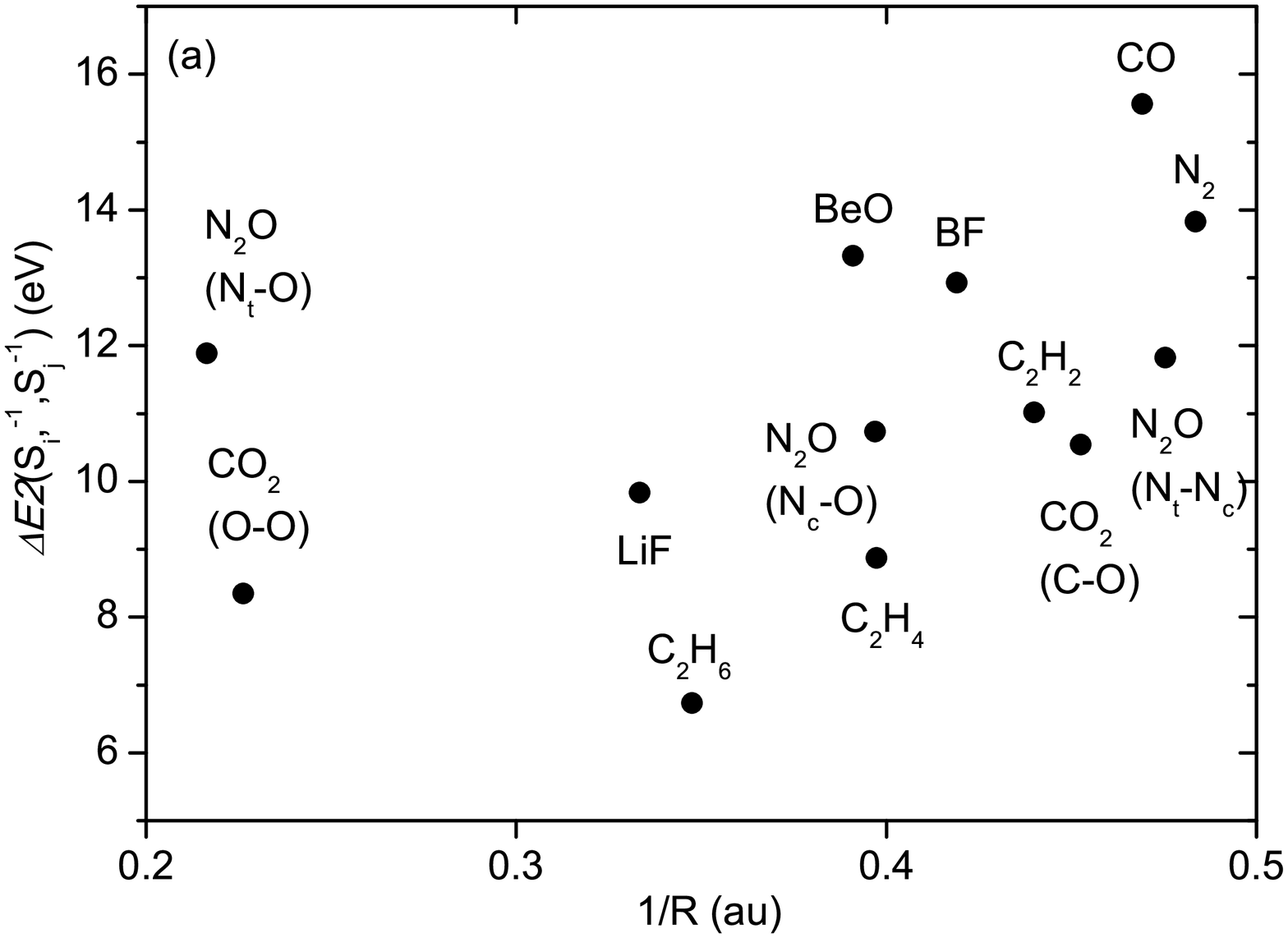}
\includegraphics[scale=0.5]{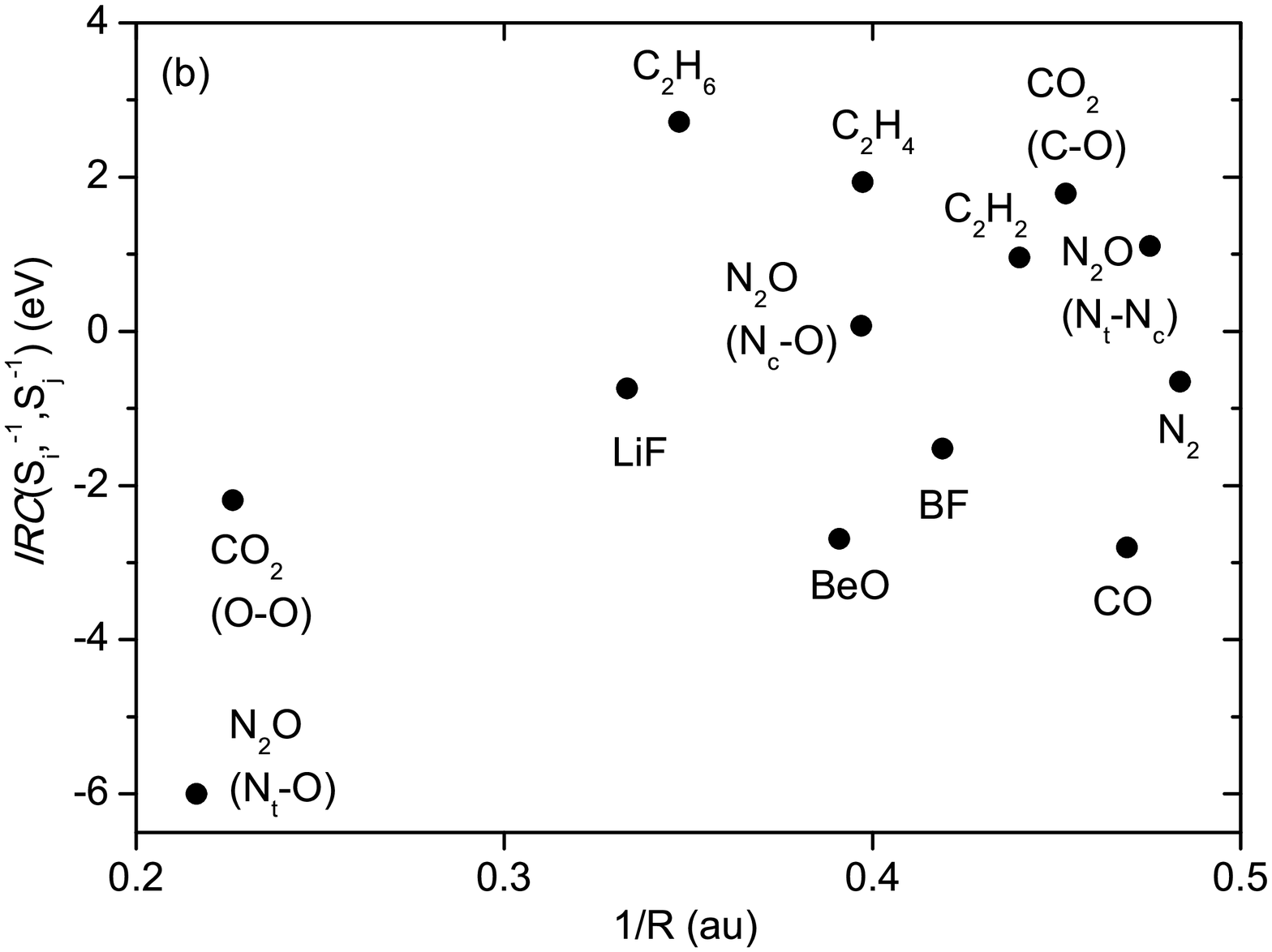}
\end{center}
\end{figure}
%$$$$$$$$$$$$$$$$$$$$$$$$$$$$$$$$$$$$$$$$$$$$$$$$$$$$$$$$$$$$$$$$$$$$

\clearpage
\newpage
\thispagestyle{empty}
%$$$$$$$$$$$$$$$$$$$$$$$$$$$$$$$$$$$$$$$$$$$$$$$$$$$$$$$$$$$$$$$$$$$$
\begin{figure}
\begin{center}
\scalebox {1.0} {\includegraphics{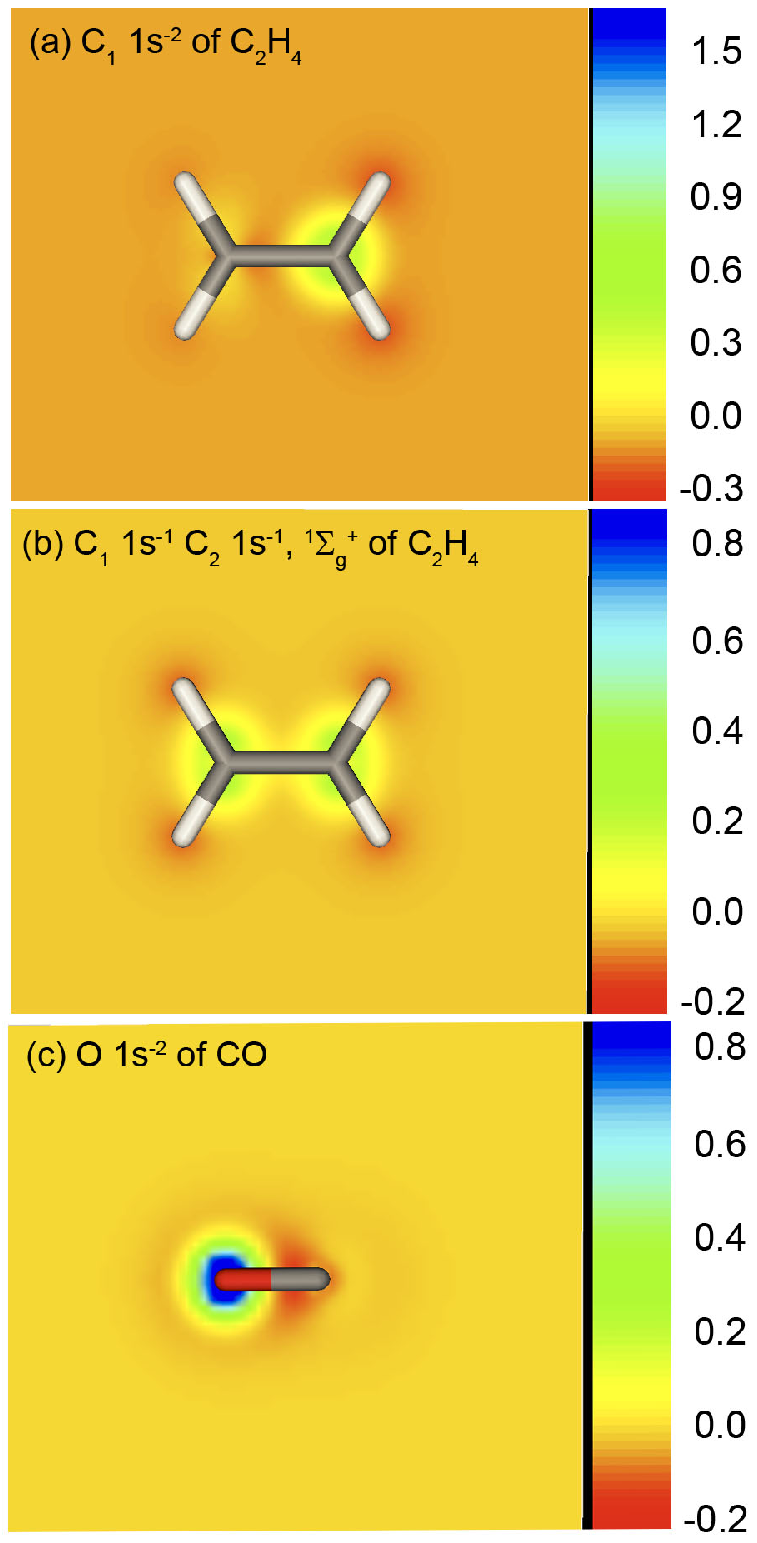}}
\end{center}
\end{figure}
%$$$$$$$$$$$$$$$$$$$$$$$$$$$$$$$$$$$$$$$$$$$$$$$$$$$$$$$$$$$$$$$$$$$$

\clearpage
\newpage
\thispagestyle{empty}
%$$$$$$$$$$$$$$$$$$$$$$$$$$$$$$$$$$$$$$$$$$$$$$$$$$$$$$$$$$$$$$$$$$$$
\begin{figure}
\begin{center}
\scalebox {0.88} {\includegraphics{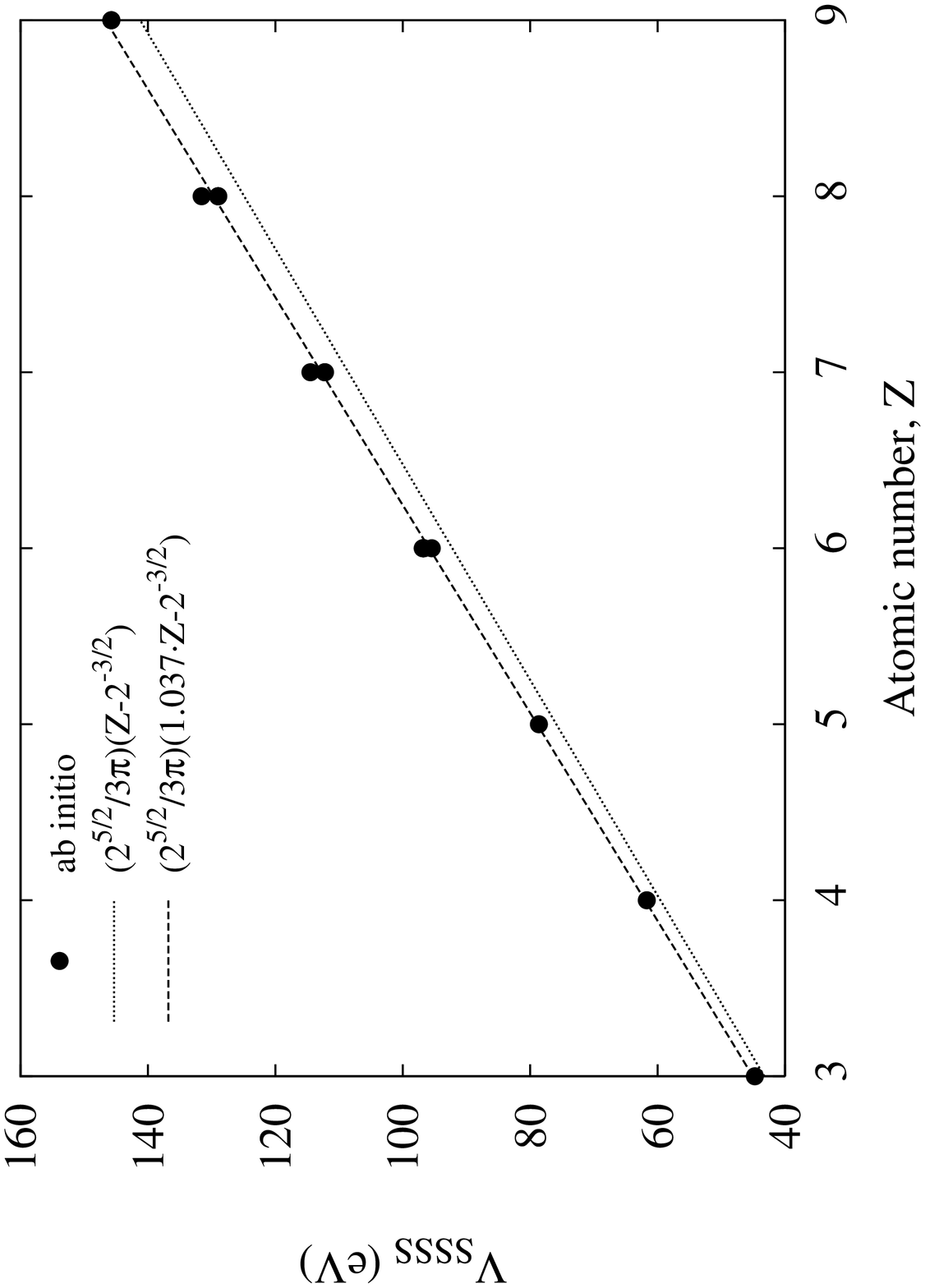}}
\end{center}
\end{figure}
%$$$$$$$$$$$$$$$$$$$$$$$$$$$$$$$$$$$$$$$$$$$$$$$$$$$$$$$$$$$$$$$$$$$$


\begin{thebibliography}{99}
\bibitem{Siegbahn71} K. Siegbahn, C. Nordling, G. Johansson, J. Hedman, P. F. Hed{\'e}n,
K. Hamrin, U. Gelius, T. Bergmark, L. O. Werme, R. Manne and Y. Baer,
{\it ESCA applied to free molecules}, North-Holland Publishing Company, Amsterdam, 1971.
\bibitem{Cederbaum86} L. S. Cederbaum, F. Tarantelli, A. Sgamellotti and J. Schirmer, J. Chem. Phys. {\bf 85}, 6513 (1986).
\bibitem{Cederbaum87_1} L. S. Cederbaum, F. Tarantelli, A. Sgamellotti and J. Schirmer, J. Chem. Phys. {\bf 86}, 2168 (1987).
\bibitem{Cederbaum87_2} L. S. Cederbaum,  Phys. Rev. A {\bf 35}, 622 (1987).
\bibitem{Ohrendorf91} E. M.-L. Ohrendorf, L. S. Cederbaum and F. Tarantelli, Phys. Rev. A {\bf 44}, 205 (1991).
\bibitem{Agren93} H. {\AA}gren, H. J{\o}rgen and Aa. Jensen, Chem. Phys. {\bf 172}, 45 (1993).
\bibitem{Reynaud96} C. Reynaud, M.-A. Gaveau, K. Bisson, Ph. Milli{\'e}, I. Nenner, S. Bodeur,
P. Archirel and B. L{\'e}vy, J. Phys. B: At. Mol. Opt. Phys. {\bf 29}, 5403 (1996).
\bibitem{Kanter06}E. P. Kanter, R. W. Dunford, B. Kr\"assig, S. H. Southworth and L. Young, 
Radiat. Phys. Chem. {\bf 75}, 2174 (2006).
\bibitem{Kheifets09} A. K. Kheifets, J.-C. Dousse, M. Berset, I. Bray, W. Cao, K. Fennane, 
Y. Kayser, M. Kavcic, J. Szlachetko and M. Szlachetko, Phys. Rev. Lett. {\bf 102}, 073006 (2009).
\bibitem{Feldhaus05} J. Feldhaus, J. Arthur and J. B. Hastings, J. Phys. B: At. Mol. Opt. Phys. {\bf 38}, S799 (2005).
\bibitem{XFEL} M. Altarelli, R. Brinkmann, M. Chergui, W. Decking, B. Dobson, S. D\"usterer, 
G. Gr\"ubel, W. Graeff, H. Graafsma, J. Hajdu, J. Marangos, J. Pfl\"uger, H. Redlin, D. Riley, 
I. Robinson, J. Rossbach, A. Schwarz, K. Tiedtke, T. Tschentscher, I. Vartaniants, H. Wabnitz, 
H. Weise, R. Wichmann, K. Witte, A. Wolf, M. Wulff, M. Yurkov, 
{\it The Technical Design Report of the European XFEL}, DESY XFEL Project Group, Deutsches 
Elektronen-Synchrotron (DESY), report No DESY 2006-097, Notkestra{\ss}e 85, 22607 Hamburg, Germany, 
July 2006.
\bibitem{SCSS} T. Shintake, H. Tanaka, T. Hara, T. Tanaka, K. Togawa, M. Yabashi, Y. Otake, 
Y. Asano, T. Bizen, T. Fukui, S. Goto, A. Higashiya, T. Hirono, N. Hosoda, T. Inagaki, S. Inoue, 
M. Ishii, Y. Kim, H. Kimura, M. Kitamura, T. Kobayashi, H. Maesaka, T. Masuda, S. Matsui, 
T. Matsushita, X. Marechal, M. Nagasono, H. Ohashi, T. Ohata, T. Ohshima, K. Onoe, K. Shirasawa, 
T. Takagi, S. Takahashi, M. Takeuchi, K. Tamasaku, R. Tanaka, Y. Tanaka, T. Tanikawa, T. Togashi, 
S. Wu, A. Yamashita, K. Yanagida, C. Zhang, H. Kitamura and T. Ishikawa, Nature Photon {\bf 2}, 
555 (2008); T. Tanaka, T. Shintake, {\it SCSS X-FEL conceptual design report}, RIKEN Harima 
Institute/SPring-8, 1-1-1, Kouto, Mikazuki, Sayo, Hyogo, Japan 679-5148, May 2005.
\bibitem{Sorokin06} A. A. Sorokin, S. V. Bobashev, K. Tiedtke and M. Richter, J. Phys. B: At. Mol. Opt. Phys. {\bf 39}, 
L299 (2006).
\bibitem{Foehlisch07} A. F\"ohlisch, M. Nagasono, M. Deppe, E. Suljoti, F. Hennies, A. Pietzsch 
and W. Wurth, Phys. Rev. A {\bf 76}, 013411 (2007).
\bibitem{Sato08} T. Sato, T. Okino, K. Yamanouchi, A. Yagishita, F. Kannari, K. Yamakawa, 
K. Midorikawa, H. Nakano, M. Yabashi, M. Nagasono and T. Ishikawa, Appl. Phys. Lett. {\bf 92}, 154103 (2008).
\bibitem{Jiang09} Y. H. Jiang, A. Rudenko, M. Kurka, K. U. K\"uhnel, T. Ergler, L. Foucar, 
M. Sch\"offler, S. Sch\"ossler, T. Havermeier, M. Smolarski, K. Cole, R. D\"orner, S. D\"usterer, 
R. Treusch, M. Gensch, C. D. Schr\"oter, R. Moshammer and J. Ullrich, Phys. Rev. Lett. {\bf 102}, 123002 (2009).
\bibitem{Fukuzawa09} H. Fukuzawa, K. Motomura, X.-J. Liu, G. Pruemper, M. Okunishi1, K. Ueda, 
N. Saito, H. Iwayama, K. Nagaya, M. Yao, M. Nagasono, A. Higashiya, M. Yabashi, T. Ishikawa, 
H. Ohashi and H. Kimura, J. Phys. B: At. Mol. Opt. Phys. 42, 181001 (2009).
\bibitem{LCLS} J. Arthur, P. Anfinrud, P. Audebert, K. Bane, I. Ben-Zvi, V. Bharadwaj, R. Bionta, 
P. Bolton, M. Borland, P. H. Bucksbaum, R. C. Cauble, J. Clendenin, M. Cornacchia, G. Decker, P. Den Hartog,
S. Dierker, D. Dowell, D. Dungan, P. Emma, I. Evans, G. Faigel, R. Falcone, W. M. Fawley, M.
Ferrario, A. S. Fisher, R. R. Freeman, J. Frisch J. Galayda, J.-C. Gauthier, S. Gierman, E. Gluskin,
W. Graves, J. Hajdu, J. Hastings, K. Hodgson, Z. Huang, R. Humphry, P. Ilinski, D. Imre, C.
Jacobsen, C.-C. Kao, K. R. Kase, K.-J. Kim, R. Kirby, J. Kirz, L. Klaisner, P. Krejcik, K.
Kulander, O. L. Landen, R. W. Lee, C. Lewis, C. Limborg, E. I. Lindau, A. Lumpkin, G. Materlik,
S. Mao, J. Miao, S. Mochrie, E. Moog, S. Milton, G. Mulhollan, K. Nelson, W. R. Nelson, R.
Neutze, A. Ng, D. Nguyen, H.-D. Nuhn, D. T. Palmer, J. M. Paterson, C. Pellegrini, S. Reiche, M.
Renner, D. Riley, C. V. Robinson, S. H. Rokni, S. J. Rose, J. Rosenzweig, R. Ruland, G. Ruocco,
D. Saenz, S. Sasaki, D. Sayre, J. Schmerge, D. Schneider, C. Schroeder, L. Serafini, F. Sette, S.
Sinha, D. van der Spoel, B. Stephenson, G. Stupakov, M. Sutton, A. Sz\"oke, R. Tatchyn, A. Toor, E.
Trakhtenberg, I. Vasserman, N. Vinokurov, X. J. Wang, D. Waltz, J. S. Wark, E. Weckert,
Wilson-Squire Group, H. Winick, M. Woodley, A. Wootton, M. Wulff, M. Xie, R. Yotam, L.
Young, A. Zewail, {\it Linac coherent light source (LCLS): Conceptual design report}, 
report No SLAC-R-593 and UC-414, April 2002.
\bibitem{Emma04} P. Emma, K. Bane, M. Cornacchia, Z. Huang, H. Schlarb, G. Stupakov and 
D. Walz, Phys. Rev. Lett. {\bf 92}, 074801 (2004).
\bibitem{Galayda} J. Galayda, (private communication).
\bibitem{Santra09} R. Santra, N. Kryzhevoi and L. S. Cederbaum, Phys. Rev. Lett. {\bf 103}, 013002 (2009).
\bibitem{Bagus65} P. S. Bagus Phys. Rev. {\bf 139} A619 (1965).
\bibitem{Werner85} H.-J. Werner and P. J. Knowles, J. Chem. Phys. {\bf 82}, 5053 (1985).
\bibitem{Hampel92} C. Hampel, K. Peterson and H.-J. Werner, Chem. Phys. Lett. {\bf 190}, 1 (1992).    
\bibitem{Dunning89} T. H. Dunning Jr., J. Chem. Phys. {\bf 90}, 1007 (1989).
\bibitem{Boys66} S. F. Boys, in {\it Quantum Theory of Atoms, Molecules and the Solid State}, edited by 
P. O. L\"owdin, (Academic Press, New York, 1966) p. 253.
\bibitem{Denis75} A. Denis, J. Langlet and J. P. Malrieu, Theor. Chim. Acta {\bf 38}, 49 (1975).
\bibitem{Cederbaum77} L. S. Cederbaum and W. Domcke, J. Chem. Phys. {\bf 66}, 5084 (1977).
\bibitem{Douglas74} M. Douglas and N. M. Kroll, Ann. Phys. {\bf 82}, 89 (1974).
\bibitem{Hess85} B. A. Hess, Phys. Rev. A {\bf 32}, 756 (1985).
\bibitem{Hess86} B. A. Hess, J. Chem. Phys. {\bf 33}, 3742 (1986).
\bibitem{Reiher04_1} M. Reiher and A. Wolf, J. Chem. Phys. {\bf 121}, 2037 (2004).
\bibitem{Reiher04_2} M. Reiher and A. Wolf, J. Chem. Phys. {\bf 121}, 10945 (2004).
\bibitem{MOLPRO} MOLPRO, version 2008, a package of {\it ab initio} programs, H.-J. Werner, 
P. J. Knowles, R. Lindh, F. R. Manby, M. Sch\"utz, P. Celani, T. Korona, A. Mitrushenkov, G. Rauhut, 
T. B. Adler, R. D. Amos, A. Bernhardsson, A. Berning, D. L. Cooper, M. J. O. Deegan, A. J. Dobbyn, 
F. Eckert, E. Goll, C. Hampel, G. Hetzer, T. Hrenar, G. Knizia, C. K\"oppl, Y. Liu, A. W. Lloyd, R. A. Mata, 
A. J. May, S. J. McNicholas, W. Meyer, M. E. Mura, A. Nicklass, P. Palmieri, K. Pfl\"uger, R. Pitzer, 
M. Reiher, U. Schumann, H. Stoll, A. J. Stone, R. Tarroni, T. Thorsteinsson, M. Wang and A. Wolf.
\bibitem{CeDoSc1980} L.S. Cederbaum, W. Domcke and J. Schirmer, Phys. Rev. A {\bf 22}, 206 (1980).
\bibitem{DaMu2009} D. Datta and D. Mukherjee, J. Chem. Phys. {\bf 131}, 044124 (2009).
\bibitem{Pickup73} B. T. Pickup and O. Goscinski, Mol. Phys.  {\bf 26}, 1013 (1973).
\bibitem{Bakke80} A. A. Bakke, A. W. Chen and W. L. Jolly, J. Electron Spectrosc. Relat. Phenom. {\bf 20}, 333 (1980).
\bibitem{Schirmer87} J. Schirmer, G. Angonoa, S. Svensson, D. Nordfors, and U. Gelius, 
J. Phys. B: At. Mol. Opt. Phys. {\bf 20}, 6031 (1987).
\bibitem{Kempgens97} B. Kempgens, H. K\"oppel, A. Kivim\"aki, M. Neeb, L. S. Cederbaum and A. M. Bradshaw, 
Phys. Rev. Lett. {\bf 79}, 3617 (1997).
\bibitem{Ehara06_1} M. Ehara, K. Kuramoto, H. Nakatsuji, M. Hoshino, T. Tanaka, M. Kitajima, H. Tanaka, 
Y. Tamenori, A. D. Fanis and K. Ueda, J. Chem. Phys. {\bf 125}, 114304 (2006).
\bibitem{Ehara07} M. Ehara, R. Tamaki, H. Nakatsuji, R. R. Lucchese, J. S\"oderstr\"om, T. Tanaka, M. Hoshino, M.
Kitajima, H. Tanaka, A. D. Fanis and K. Ueda, Chem. Phys. Lett. {\bf 438}, 14 (2007).
\bibitem{Ehara06_2} M. Ehara, H. Nakatsuji, M. Matsumoto, T. Hatamoto, X.-J. Liu, T. Lischke, 
G. Pruemper, T. Tanaka, C. Makochekanwa, M. Hoshino, H. Tanaka, J. R. Harries, Y. Tamenori, K. Ueda, 
J. Chem. Phys. {\bf 124}, 124311 (2006).
\bibitem{Hatamoto07} T. Hatamoto, M. Matsumoto, X.-J. Liu, K. Ueda, M. Hoshino, K. Nakagawa, T. Tanaka, H. Tanaka, M.
Ehara, R. Tamaki and H. Nakatsuji, J. Electron Spectrosc. Relat. Phenom. {\bf 155}, 54 (2007).
\bibitem{Angonoa87} G. Angonoa, O. Walter and J. Schirmer, J. Chem. Phys. {\bf 87}, 6789 (1987). 
\bibitem{Ehara09} M. Ehara, K. Kuramoto and H. Nakatsuji, Chem. Phys. {\bf 356}, 195 (2009).
\end{thebibliography}
\end{document}